\documentclass[aps,prb,twocolumn,showpacs,superscriptaddress]{revtex4-1}

\usepackage{color}
\usepackage[nottoc,numbib]{tocbibind}
\usepackage{bm}
\usepackage{amsmath,amsfonts,amstext,amssymb}
\usepackage{graphicx}

\usepackage{ifthen}

\newcommand{\mZ}{\mathcal{Z}}
\newcommand{\mC}{\mathcal{C}}
\newcommand{\mH}{\mathcal{H}}
\newcommand{\mR}{\mathcal{R}}
\newcommand{\mO}{\mathcal{O}}
\newcommand{\mN}{\mathcal{N}}
\newcommand{\thb}{\tanh(\beta)}

\newcommand{\be}{\begin{equation}}
\newcommand{\ee}{\end{equation}}
\newcommand{\bea}{\begin{eqnarray}}
\newcommand{\eea}{\end{eqnarray}}

\begin{document}

\setcitestyle{square}
 
\title{The Random-Diluted Triangular Plaquette Model: \\ study of phase transitions in a Kinetically Constrained Model}

\author{Silvio Franz}
\affiliation{LPTMS, CNRS, Universit\'e Paris Sud et Universit\'e Paris-Saclay, 91405 Orsay, France}

\author{Giacomo Gradenigo}
\email{ggradenigo@gmail.com}
\affiliation{LPTMS, CNRS, Universit\'e Paris Sud et Universit\'e Paris-Saclay, 91405 Orsay, France}
\affiliation{IPhT, CEA Saclay, F-91191 Gif-sur-Yvette Cedex, France}
\affiliation{Universit\'e Grenoble Alpes, LIPHY, F-38000 Grenoble, France \\ CNRS, LIPHY, F-38000 Grenoble, France}

\author{Stefano Spigler}
\affiliation{LPTMS, CNRS, Universit\'e Paris Sud et Universit\'e Paris-Saclay, 91405 Orsay, France}

\date{\today}

\begin{abstract}

  We study how the thermodynamic properties of the Triangular
  Plaquette Model (TPM) are influenced by the addition of extra
  interactions. The thermodynamics of the original TPM is trivial,
  while its dynamics is glassy, as usual in Kinetically Constrained
  Models.  As soon as we generalize the model to include additional
  interactions, a thermodynamic phase transition appears in the
  system.  The additional interactions we consider are either short
  ranged, forming a regular lattice in the plane, or long ranged of
  the small-world kind. In the case of long-range interactions we call
  the new model Random-Diluted TPM. We provide arguments that the
  model so modified should undergo a thermodynamic phase transition,
  and that in the long-range case this is a glass transition of the
  ``Random First-Order'' kind. Finally, we give support to our
  conjectures studying the finite temperature phase diagram of the
  Random-Diluted TPM in the Bethe approximation.  This correspond to
  the exact calculation on the random regular graph, where free-energy
  and configurational entropy can be computed by means of the cavity
  equations.

\end{abstract}

\maketitle 



\section{Introduction} 
\label{S0}
There are two main points of view to understand the nature of the
glass phase. On one hand there is the idea of ``dynamic
facilitation'', which emphasizes the role of frustration in the
dynamics: the motion of the microscopic constituents of a
glass-forming system becomes inhibited by their close neighbours when
the system is cooled. The consequence is that the material remains
stuck for an extremely long time in a certain amorphous
configuration. On the other hand there is the landscape scenario, or
``Random First-Order Transition'' (RFOT)
theory~\cite{KTW89,MP00,LW07,PZ10}, according to which the formation
of a glass is the reflex of the existence of metastable states whose
multiplicity is strongly reduced as the temperature is lowered.  In
the dynamic facilitation scenario thermodynamics is deemphasized and
the interaction only plays a role in dynamics. This idea is at the
basis of the description of glasses provided by Kinetically
Constrained Models (KCM)~\cite{gst11,FA84,ej91,KA93,LJ00,YC02}. KCM
are lattice models where the variables do not have energetic
interactions, but are subject to dynamic constraints. Both theories
have reference models that reproduce important aspects of glass
phenomenology, and it is hard to decide which scenario is the
appropriate one to describe real systems. The two descriptions are
very similar at the mean-field level: recent numerical simulations
have shown that both at the level of average behavior~\cite{SEL13} and
at the level of fluctuations~\cite{SELFRANZ13} KCM models on random
graphs~\cite{SBT05} follow a glass transition pattern predicted by
Mode Coupling Theory. In~\cite{FKZ12} it was also shown that a typical
RFOT model, the XOR-SAT on random graph, can be mapped in a KCM. The
XOR-SAT model combines salient features of both theories: it has a
trivial high-temperature thermodynamics as in KCM, but with a finite
temperature entropy crisis glass transition as in RFOT. It is an
interesting question if this commonality of mechanisms observed in
mean-field extends to finite dimensions.

To investigate this question, in the present work we analyze the
thermodynamic properties of a modified 2D Triangular Plaquette Model
(TPM)~\cite{nm99,gn00,jbg05,jg05}. The original TPM is an example of
KCMs: it is a spin model whose thermodynamics is the one of a trivial
paramagnet while at the same time the model displays dynamical glassy
phenomena with a super-Arrhenius relaxation time. Indeed, the TPM is
nothing but a realization of the XOR-SAT model in finite (two)
dimensions~\cite{FKZ12}. Our attention was brought to the TPM in
particular by the results of~\cite{G14}. In~\cite{G14}, and more
recently in~\cite{TJG15}, it has been shown that the TPM, in presence
of external fields, supports both \emph{dynamic} and
\emph{thermodynamic} phase transitions to glassy arrested
phases. Emphasis is put in~\cite{G14,TJG15} on the fact that as soon
as such external fields are switched off ergodicity is restored, so
that the TPM is just \emph{marginally} glassy. Looking at the results
of of~\cite{G14,TJG15} from another perspective, they provide an
evidence that the trivial thermodynamics of the TPM can be
dramatically altered by means of very small perturbations. Following
this line, our purpose here is it to show that the triviality of the
TPM thermodynamics is marginal and its physics is close to the one of
the landscape scenario. We will show that, as soon as some new
interactions are introduced in the TPM, its thermodynamics cannot be
trivial anymore and a phase transition appear in the system. Our
results support the point of view that the dynamic facilitation and
the landscape scenario should be regarded as complementary rather than
alternative.

Specifically, the TPM is a lattice spin model where the spins sit on a
two-dimensional triangular lattice endowed with ``plaquette''
interactions: each plaquette corresponds to the product of the three
spins placed at the corners of an upward triangular cell of the
lattice. The energy of the model is $\mathcal{H} = -\sum_{\langle ijk
  \rangle} \sigma_i\sigma_j\sigma_k$, where each triplet $\langle ijk
\rangle$ of indices is associated to a plaquette. The remarkable
observation made in~\cite{nm99} is that there is a one-to-one
correspondence between spins and plaquettes: a configuration of the
system is well defined either assigning the values of the spins
$\{\sigma_i\}_{i=1,\ldots,N}$ or the values of the plaquette variables
$\{\tau_a\}_{a=1,\ldots,N}$, where $\tau_a =
\sigma_{i(a)}\sigma_{j(a)}\sigma_{k(a)}$. In terms of the plaquette
variables the Hamiltonian is equivalent to that of a system of
non-interacting spins in a field, for which the partition function can
be trivially calculated: $\mathcal{Z}=2^N [\cosh(\beta)]^N$. The
absence of thermodynamic singularities, together with a critical
slowing down at low temperatures~\cite{nm99,gn00},
$\tau_{\textrm{rel}}\sim \exp(A/T^2)$, are typical features of
KCMs. If one performs local MC updates acting upon the spins but then
looks at the resulting dynamics of the plaquette variables, the latter
looks like a kinetically constrained dynamics. It happens that the
annihilation of a ``defect'', namely the flip of a plaquette ``$a$''
from $\tau_a=-1$ to $\tau_a=1$, is favoured only when there is
\emph{at least} one other defect, i.e. excited plaquette, connected to
``$a$''. Two plaquettes are connected when they have a spin in
common. Since the dynamics acts upon spins, and in the TPM the update
of one spin always corresponds to the updated of three plaquettes ,
the Monte Carlo dynamics corresponds to the flipping of three
plaquettes per time. Due to the odd number of plaquettes attached to a
spin $\sigma_i$, in the TPM there is no spin flip with $\Delta E=0$,
since $E=-\sum_{a\in \partial i}\tau_a$. For the same reason one needs
that at least two of the plaquettes attached to $\sigma_i$ are
excited, namely they are both $\tau_a=-1$, in order to have $\Delta
E<0$ by flipping $\sigma_i$. This explains why the annihilation of a
defect is favoured only in the vicinity of another defect. The
transition rates for spin updates depend on temperature through the
standard metropolis rule, namely each attempted update is accepted
with probability $p=\textrm{min}(1,e^{-\beta \Delta E})$. The idea we
want to test in this paper is that if we introduce some new
interactions between variables (or if we remove some), an important
parameter in the thermodynamics is the ratio $\alpha=M/N$ between the
number $M$ of plaquettes in the Hamiltonian and the number $N$ of
spins in the systems. While such a scenario is well established for
plaquette models on random graphs~\cite{mrz03}, this has not been
tested to our knowledge in finite dimensional geometries. Of course
the ratio $\alpha$ can be changed in many different ways. Here we will
focus on two class of models with extra plaquettes: in a first class
we choose the triplets of spin in the new interactions completely at
random, in a second class the new interactions are taken on a regular
sublattice of the triangular lattice, either a fraction of them chosen
randomly or all the interactions of the sublattice. With the first
choice we induce arbitrarily long-range interactions, so that the
resulting model is a kind of small-word network~\cite{bw00}. In order
to be general, in the model with long-range interactions we also take
into account a dilution of the plaquettes of the original
two-dimensional triangular lattice: that is why we call such a model
the Random-Diluted Triangular Plaquette Model.  The paper is
structured as follows. In the first part, Sec.~\ref{S1}, we present
the two class of modified TPMs just mentioned. For each class we
discuss the behaviour of the high-temperature expansion and present
the results of numerical simulations.
In Sec.~\ref{S2} we discuss how to use the leaf-removal
algorithm~\cite{mrz03}, which is a method borrowed from the study of
constrained optimization problems, to draw a \emph{tentative} phase
diagram at $T=0$ of the Random-Diluted TPM in the
$(\alpha_s,\alpha_L)$ plane: $\alpha_L$ is the concentration of
long-range plaquettes while $\alpha_s$ is the concentration of
short-range plaquettes in the two-dimensional lattice. Finally, in
Sec.~\ref{S3} of the paper we present the phase diagram at finite
temperature for the Random-Diluted TPM on the random regular graph,
which we will refer to often in the paper also as the Bethe lattice
geometry, where the glass transition temperature can be exactly
calculated by solving the cavity equations.

\section{Triangular Plaquette Model with additional interactions: high-temperature expansion and numerical simulations}
\label{S1}

The simplest choice of additional interactions for a modified TPM with
Hamiltonian $H=H_{\textrm{TPM}}+H_{\textrm{extra}}$ is represented by
new ferromagnetic plaquettes: $H_{\textrm{extra}}=-\sum_{ijk} \sigma_i
\sigma_j \sigma_k$. In the Hamiltonian $H_{\textrm{extra}}$ the only
source of randomness is then represented by the choice of which spins
participate in each of the new plaquettes, i.e. the choice of the
triplets of indices $ijk$. The ground state of the so modified TPMs is
the configuration where all spins attain the value $\sigma_i=+1$. The
aim of this work is to provide an evidence that in some of these
modified TPMs a glass transition takes place before the system has the
time to relax to the ordered ground state. We already know from the
literature that this is the case for the TPM on a random
graph~\cite{fmrwz01}. The idea of introducing new interactions is
motivated by the purpose to induce an entropic crisis in the system,
which is the typical mechanism for the formation of a glass phase
within the RFOT theory scenario~\cite{AC09}. We have already mentioned
that in the TPM the partition function is simply $\mathcal{Z}=2^N
[\cosh(\beta)]^N$. Let us assume the existence of a modified TPM such
that the number of plaquettes $M$ is different from the number of
spins $N$, but the partition function is still the trivial one:
$\mathcal{Z}=2^N [\cosh(\beta)]^M$. In this hypothetical TPM the
entropy of the system would be \be s(\beta)=\log(2)+\frac{M}{N}
\log(\cosh(\beta))-\frac{M}{N}\beta\tanh(\beta) \label{eq:ZTPM0} \ee
which in the limit of zero temperature yields \be
\lim_{\beta\rightarrow\infty} s(\beta) = \left(1 - \frac{M}{N} \right)
\log(2) \label{eq:ZTPM} \ee Eq.~(\ref{eq:ZTPM}) tells us that in the
case when $\alpha=M/N>1$ the entropy is negative at $T=0$, so that an
entropic crisis takes place at $T>0$. Clearly, since the TPM and any
decoration of it are models with discrete variables, any expression
yielding a negative entropy cannot be an exact one. The only
possibility is that, in a TPM with additional plaquettes ($M/N>1$),
the expression of the entropy in Eq.~(\ref{eq:ZTPM0}) is an
approximated one, for instance it can be the one provided by
high-temperature expansion. This expansion usually well describes the
properties of the liquid phase: the finding of a negative entropy by
its prolongation to low temperatures tells us that the description of
the system as in the liquid state becomes inconsistent and the glass
comes into play~\cite{AC09}. Our purpose here is to show that, in some
modified TPMs, the parameter which controls the entropy crisis in the
high-temperature phase is the ratio $\alpha=M/N$ between the number of
plaquettes and the number of spins. While it is well know for
mean-field plaquette models that the glass transition is controlled by
such a ratio $\alpha=M/N$~\cite{mrz03}, we provide here some evidences
that this should be the case also in finite dimensions.  All the TPMs
with additional interactions we discuss here are characterized by an
Hamiltonian of the kind
\begin{equation}
 H = H_{\textrm{TPM}} + H_{\textrm{extra}},
\label{eq:modello}
\end{equation}
where two different contributions read off: $H_{\textrm{extra}} = -
\sum_{r=1}^{M_L} \sigma_{i_r} \sigma_{j_r} \sigma_{k_r}$, with the
index $r$ running over the new additional plaquettes, and the standard
Hamiltonian of the TPM model, $H_{\textrm{TPM}} = - \sum_{s=1}^{M_s}
\sigma_{i_s} \sigma_{j_s} \sigma_{k_s}$, with the index $s$ running
over the plaquettes of the two-dimensional triangular lattice. The two
parameters which characterize the model are the concentrations of
original TPM plaquettes and of new additional plaquettes: respectively
$\alpha_s=M_s/N$ and $\alpha_L=M_L/N$. The models with additional
plaquettes connecting arbitrarily far apart spins and those with
additional short-range plaquettes will be presented respectively in
Sec~\ref{S1a} and Sec.~\ref{S1b}. In Sec~\ref{S1a} and Sec.~\ref{S1b}
we study the high-temperature expansion and present some numerical
results for the simplest case in which $\alpha_L$ is arbitrary but
$\alpha_s$ is fixed to $\alpha_s=1$, that is when there is no dilution
of the plaquettes in the original 2D triangular lattice. Since the
analysis of subsections~Sec.\ref{S1a} and~Sec.\ref{S1b} shows that the
best candidate to display a low temperature glass transition is the
TPM model with additional long-range plaquettes (``small-word''
network), this model will be studied in the rest of the paper
(Sec.~\ref{S2} and Sec.~\ref{S3}) also taking into account values
$\alpha_s<1$. The model in which both $\alpha_s$ and $\alpha_L$ take
values in the interval $[0,1]$ is the Random-Diluted TPM.

\subsection{Long-range additional plaquettes: the Random-Diluted TPM}

\label{S1a}
A generalized TPM with additional plaquettes has the following 
high-temperature expansion for the partition function:
\be \mZ = 2^N
\left[\cosh(\beta)\right]^M \left( 1 + \sum_{m=1}^{M} \mC(m,M)
     [\tanh(\beta)]^m \right),
\label{eq:ZGEN}
\ee where the sum on the right hand side of Eq.~(\ref{eq:ZGEN}) runs
over \emph{hyperloops}~\cite{RWZ01} made of $m$ plaquettes: $\mC(m,M)$
represents the number of hyperloops of $m$ plaquettes that can be
built in a system with a total number $M$ of plaquettes. An hyperloop
is defined~\cite{RWZ01} as a set of interactions such that each spin
is appearing an even number of times. The precise expression of
$\mC(m,M)$ depends on the model. For instance, in the pure TPM, one
has $\mC(m,M=N)=0$ and the high-temperature series yields exactly the
partition function. Other situations will be discussed in the
following paragraphs. Let us focus here on the behaviour of the
high-temperature expansion in the case when the $M_L=\alpha_L N$
plaquettes of $H_{\textrm{extra}}$ have spins drawn with uniform
probability from the lattice, with the only constraint that spins in
the same plaquette are different. Let us indicate with $\mathcal{R}$
an instance of disorder, namely a particular choice of the triplets of
spins in the plaquettes of $H_{\textrm{extra}}$. It is also convenient
to introduce the average of a function $f[\boldsymbol \sigma]$, with
$\boldsymbol \sigma$ denoting a configuration of the system
$\boldsymbol \sigma = (\sigma_1,\ldots,\sigma_N)$, with respect to the
measure provided by the pure TPM: \be \langle f[\boldsymbol \sigma]
\rangle_{ _{\textrm{TPM}}} = \frac{1}{\mathcal{Z}_{TPM}}
\sum_{\boldsymbol \sigma} e^{-\beta H_{\textrm{TPM}}[\boldsymbol
    \sigma]} f[\boldsymbol \sigma] \ee The partition sum for a given
$\mathcal{R}$ reads then:
\begin{eqnarray}
\mathcal{Z}_{\mathcal{R}} &=& \sum_{\sigma} e^{ - \beta H_{\textrm{TPM}}[\boldsymbol \sigma]+ \beta \sum_{r}^{M_L}
  \sigma_{i_r} \sigma_{j_r} \sigma_{k_r},} \nonumber \\ &=&
        2^N [\cosh(\beta)]^{N+M_L} \left\langle \prod_r (1 + \sigma_{i_r}\sigma_{j_r} \sigma_{k_r} \thb) \right\rangle_{ _{\textrm{TPM}}} \nonumber \\ 
&=& 2^N [\cosh(\beta)]^{N+M_L} \Big[ 1+ \sum_{r \in \mathcal{R}}^{M_L} \langle \tau_r \rangle_{ _{\textrm{TPM}}} \thb + \nonumber \\ &+&
          \sum_{r\neq p \in \mathcal{R}}^{M_L} \langle \tau_r \tau_p
          \rangle_{ _{\textrm{TPM}}} [\thb]^2 + \ldots \Big]
\label{eq:high-temp}
\end{eqnarray}
where, to lighten the notation, we used the plaquette variables
$\tau_r=\sigma_{i_r}\sigma_{j_r} \sigma_{k_r}$ and $M_L$ is the number
of additional plaquettes in $H_{\textrm{extra}}$.  From
Eq.~(\ref{eq:high-temp}) we see that the product of $m$ ``long-range''
plaquettes contributes an hyperloop in the high-temperature expansion
\emph{if and only if} in the original TPM the corresponding
correlation function $\langle \sigma_{i_1}\sigma_{j_1}
\sigma_{k_1}\ldots \sigma_{i_m}\sigma_{j_m} \sigma_{k_m}\rangle_{
  _{\textrm{TPM}}}$ is finite in the thermodynamic limit. Let us now
notice that in our modified TPM with long-range plaquettes the
annealed and quenched averages are equivalent at high temperature due
to the fact that the partition function is self-averaging,
i.e. $\overline{\mathcal{Z}}^2=\overline{\mathcal{Z}^2}$: the proof of
self-averaging is presented in Appendix A. This equivalence of
annealed and quenched averages in the high-temperature phase is
typical not only of the the Random Energy Model~\cite{derrida}, the
simplest model with an entropy-driven glass transition like the one we
expect in our modified TPM, but also of all the p-spin models
fully-connected or on a random (hyper-)graph. In all these situations
the behaviour of the thermodynamic potentials in the high-temperature
phase is obtained from the annealed free-energy $f = - (\beta
N)\log(\overline{\mathcal{Z}})$. We are therefore allowed to average
over disorder the partition function:
\bea 
&\overline{\mathcal{Z}}& = \frac{1}{\mN(\mR_M)} \sum_{\mR_M} \mathcal{Z}_{\mathcal{R}} = \nonumber \\ 
&=& 2^N [\cosh(\beta)]^{N+M_L} \bigg[ 1 + \frac{\binom{M_L}{1}}{\mN(\mR_1)} \sum_{r=1}^{\mN(\mR_1)}
          \langle \tau_r \rangle_{ _{\textrm{TPM}}} \thb + \nonumber
          \\ &+& \frac{\binom{M_L}{2}}{\mN(\mR_2)} \sum_{r,p=1}^{\mN(\mR_1)} 
             \langle \tau_r \tau_p \rangle_{_{\textrm{TPM}}} [\thb]^2 + \dots \bigg] \nonumber \\ 
\label{eq:partition1}
\eea
In Eq.~(\ref{eq:partition1}) the indices $r$, $p$ run over the set of
\emph{all} possible choices of random triplets of spins, while
$\mN(\mR_k)$ is the number of ways in which $k$ random triplets can be
chosen, which, in the limit $N \gg k \gg 1$, reads \be \mN(\mR_k) =
\binom{N}{3}^k \sim \mO(N^{3k}).\ee In each sum on the right-hand side
of Eq.~(\ref{eq:partition1}) the number of choices for the plaquettes
not appearing in the brackets $\langle~\rangle_{ _{\textrm{TPM}}}$ has
been canceled out with the corresponding factor in the normalization
constant. The binomial coefficient in front of each summation symbol
in Eq.~\ref{eq:partition1} represents the number of ways to choose $m$
plaquettes out of $M_L$. Taking then into account that the typical
value $\langle \tau_r \tau_p\tau_q\rangle_{ _{\textrm{TPM}}}$ does not
depend on the choice of the indices, the expression in
Eq.~(\ref{eq:partition1}) simplifies to
\bea
\mZ &=& 2^N [\cosh(\beta)]^{N+M_L} \cdot \nonumber \\ 
&&\left[ 1 + \sum_{m=1}^{M_L} \binom{M_L}{m} \langle \tau_1\ldots\tau_m \rangle_{ _{\textrm{TPM}}} [\thb]^m \right]
\label{eq:hightemp0}
\eea
It can be then proven, as shown in Appendix A, that 
\be \lim_{N\rightarrow\infty} \binom{M_L}{m} \langle \tau_1 \ldots \tau_m
\rangle_{ _{\textrm{TPM}}} = 0,\ee so that the high-temperature expansion is 
trivial and we have 
\begin{equation}
\overline{\mathcal{Z}} = 2^N [\cosh(\beta)]^{N+M_L}. 
\label{eq:hightemp}
\end{equation}
The consequence of Eq.~(\ref{eq:hightemp}) is that, since $\alpha =
(N+M_L)/N > 1$, according to the expression in Eq.~(\ref{eq:ZTPM}), at
$T=0$ the entropy is negative, which in turn implies an entropic
crisis at $T>0$.  As already mentioned, the finding of negative
entropy is, for a model like ours with discrete variables, an
artifact of the high-temperature expansion: a phase transition usually
takes place preventing this to happen. Such a phase transition is not
necessary a glass one: indeed we find from the numerical simulations
discussed below that a first-order transition to an ordered state is
taking place. As shown in~\cite{fmrwz01} for a TPM on
random graph, the presence of an ordered ground state does not spoil
the possibility of a glass transition.\\

\begin{figure}[b!]
\includegraphics[width=\columnwidth]{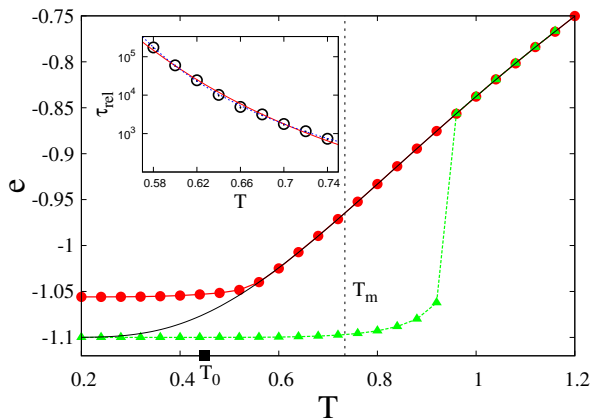}
\caption{(Color online) \emph{Main}: Energy hysteresis cycle for a TPM
  model with additional long-range plaquettes. We simulated the model
  on a lattice with the shape of a rhombus (see for instance the
  representation of Fig.~\ref{fig:ITPM_lattice}) with $L=128$
  plaquettes per side and a concentration of additional plaquettes
  $\alpha_L=0.1$.  (Red) Circles: cooling (cooling rate from $10^6$ to
  $10^7$ MCS per $\Delta T=0.04$); (Green) Triangles: heating;
  Continuous (black) line: high-temperature expansion,
  $e(T)=-(\alpha_L+1)\tanh(1/T)$. The melting temperature $T_m$ is
  indicated by the dotted vertical line while the temperature $T_0$ of
  the entropy crisis of the high-temperature expansion is the filled
  (black) square. \emph{Inset}: Circles are relaxation time in the
  supercooled liquid phase as function of $T$, and lines are
  respectively the two parameters fit with $a\exp(b/T)$ (continuous
  red line) and the three parameters fit with $c\exp(d/(T-T_K))$. The
  critical temperature obtained from the fit is $T_K=0.374$.}
\label{fig:hysteresis-LR}
\end{figure}

\paragraph*{Numerical Simulations} In order to test the behaviour of
the TPM in presence of extra plaquettes with randomly chosen spins we
realized numerical simulations of the model. We studied a lattice with
triangular cells and the shape of a rhombus with $L=128$ plaquettes
per side, periodic boundary conditions, and a concentration of extra
random plaquettes $\alpha_L = 0.1$. In order to speed up the
equilibration dynamics at low temperatures we used a rejection-free
algorithm~\cite{BKL75}. The behaviour of the internal energy along an
hysteresis cycle is represented by data in
Fig.~\ref{fig:hysteresis-LR}.  The scenario is as follows: by cooling
down the system from high temperatures one first finds a spinodal
temperature $T_s$ where an ordered metastable state appears. By
further cooling the systems it is then found a melting temperature
$T_m$ where the ordered state becomes stable. The melting temperature
has been determined from the free-energy obtained numerically via
thermodynamic integration: $f(\beta_1) = \beta_0 \beta_1^{-1}
f(\beta_0) + \beta_1^{-1} \int_{\beta_0}^{\beta_1} d\beta
e(\beta)$. In panel a) of Fig.~\ref{fig:hysteresis-LR} there are two
data sets which correspond respectively to the paramagnetic phase, at
the higher energy $e^{\textrm{para}}(\beta)$, and the ordered phase,
at the lower energy $e^{\textrm{ferro}}(\beta)$: these two phases are
stable respectively above and below the melting temperature $T_m$. By
independently performing the thermodynamic integration over
$e^{\textrm{para}}(\beta)$ and $e^{\textrm{ferro}}(\beta)$ one gets
respectively $f^{\textrm{para}}(\beta)$ and
$f^{\textrm{ferro}}(\beta)$: $T_m$ is determined by the crossing of
$f^{\textrm{para}}(\beta)$ and $f^{\textrm{ferro}}(\beta)$.
Since for $T>1$ data are well interpolated by the high-temperature
expansion, we fix the integration constant $f(\beta_0)$ to $f(\beta_0)
=-\beta_0\log(2)-\beta_0(1+\alpha_L)\log[\cosh(\beta_0)]$, with
$\beta_0^{-1}=1.2$. In the top panel of Fig.~\ref{fig:hysteresis-LR},
the melting temperature $T_m$ is represented by a vertical dotted
line: it is interesting to note that the system can be cooled to a
remarkable extent below $T_m$ while remaining in the disordered liquid
phase. Bringing $T$ down further, the relaxation time becomes so large
that the system falls out of equilibrium. Our cooling protocol is
represented by runs ranging from $10^6$ to $10^7$ Monte Carlo sweeps
for each temperature, and by temperature jumps of $\Delta T =
0.04$. Using this protocol we were not able to detect any tendency of
the system to relax to the ordered ground state. From simulations we
learn therefore that a first-order transition is present at $T_m$, but
the systems is highly stable in the supercooled liquid-phase, namely
at temperatures $T<T_m$. This finding suggests that the system avoids
on the time scales we sampled the negative entropy obtained by
extrapolating the high-temperature expansion just by forming a
glass. Let us also note that the annealed high-temperature expansion
in Eq.~(\ref{eq:hightemp}) has a very good agreement with simulations:
in Fig.~\ref{fig:hysteresis-LR} it can be seen that within the whole
range of temperatures where we could equilibrate the system no
relevant departure of data from the annealed energy $e(\beta)=-
(1+\alpha_L) \tanh(\beta)$ can be detected. In
Fig.~\ref{fig:hysteresis-LR} we can also see that the temperature
$T_0$ of the entropy crisis, as can be estimated from the
high-temperature expansion of the entropy in Eq.~(\ref{eq:ZTPM0}),
lies below the temperature range where we can equilibrate the
system. With respect to the ideal glass transition temperature $T_K$,
the temperature $T_0$ can be regarded as a lower bound for the
possible values of $T_K$. It is hard to prove or disprove the
existence of an ideal glass transition for our modified TPM solely on
the basis of numerical data: this is clear looking, for instance, at
the inset of the top panel of Fig.~\ref{fig:hysteresis-LR} where the
equilibrium relaxation time $\tau_{\textrm{rel}}$ is shown as a
function of temperature. Good fits of the data (see the inset of the
top panel in Fig.~\ref{fig:hysteresis-LR}) can be obtained either with
a function diverging at finite temperature, $\tau_{\textrm{rel}} \sim
\exp(A/(T-T_K))$, or with the super-Arrhenius law $\tau_{\textrm{rel}}
\sim \exp(B/T^2)$ (where clearly $A \neq B$): this is an ambiguous
situation that is well know for these kind of models~\cite{afr96} and
in general for glasses. That is why, in order to have a better insight
into the thermodynamics of the TPM with random plaquettes, we propose
in Sec.~\ref{S2} a different approach based on tools and ideas
borrowed from the study of constraint satisfaction problems.
\vspace{0.7cm}
\begin{figure}[h!]
\centering
\includegraphics[width=\columnwidth]{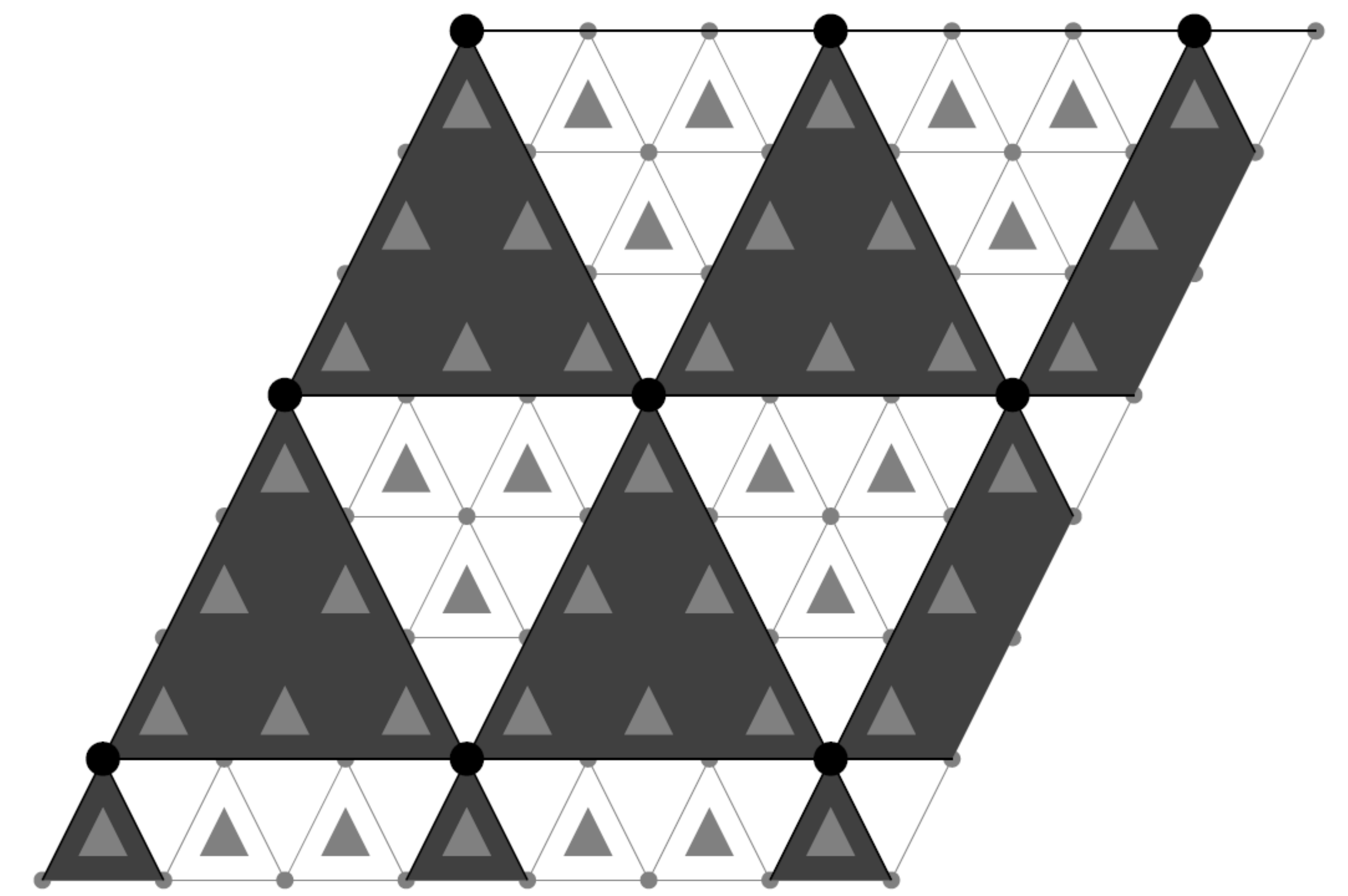}
\caption{(Color online) \small Triangular lattice for the generalized TPM with
  additional short-range plaquettes on a regular sublattice made of
  equilateral triangles of side 3. The original TPMs plaquettes and
  additional interactions are shown as light gray and dark gray
  triangles, respectively.}
\label{fig:ITPM_lattice}
\end{figure}
\subsection{Short-range additional plaquettes}
\label{S1b}
In the previous section we studied the effect of long-range random
interactions on the high-temperature expansion of the triangular
plaquette model.  It is then interesting to investigate what happens
in presence of short-range interactions. Our choice for the new kind
of short-range interactions has been guided by the purpose to reduce
as much as possible the corrections to the high-temperature
expansion. Since in the TPM the correlation of any group of three
spins placed at the corners of an equilateral triangle of side $3$ is
zero, we defined an additional sublattice formed by equilateral
triangular cells of that kind. The additional plaquettes appearing in
$H_{\textrm{extra}}$ are then chosen as all the plaquettes of this
regular sublattice.  In this case we know that at least the first term
of the high-temperature expansion vanishes. A representation of this
\emph{odd} sublattice is given in Fig.~\ref{fig:ITPM_lattice}, where
the additional plaquettes correspond to the spins placed at the
corners of equilateral triangles with side 3. Let us then rewrite the
partition function as a series in $\tanh(\beta)$:
\bea \mathcal{Z}
  &=& 2^N[\cosh(\beta)]^{N+M_L} \bigg[ 1+ \sum_r^{M_L} \langle \tau_r
    \rangle_{ _{\textrm{TPM}}} \thb + \nonumber \\ &+& \sum_{r\neq
      p}^{M_L} \langle \tau_r \tau_p \rangle_{ _{\textrm{TPM}}}
           [\thb]^2 + \ldots \bigg],
\label{eq:new-partition}
\eea
where the sums in Eq.~(\ref{eq:new-partition}) run over all the
plaquettes of the odd sublattice. At this stage the expression in
Eq.~(\ref{eq:new-partition}) is exact: it is just a way of rewriting
the partition function. The difference between
Eq.~(\ref{eq:new-partition}) above and Eq.~(\ref{eq:partition1}) in
the previous section is that in Eq.~(\ref{eq:new-partition}) each term
of the kind
\be \sum_{r_1,\ldots,r_k}^{M_L} \langle \tau_{r_1} \ldots
\tau_{r_k} \rangle_{ _{\textrm{TPM}}} [\thb]^k, 
\ee 
accounts for the hyperloops made with all the possible choices of $k$
plaquettes from the additional sublattice, while in
Eq.~(\ref{eq:partition1}) only the hyperloops coming from a given
random choice of the additional plaquettes, denoted with
$\mathcal{R}$, were taken into account.  Due to the regular structure
of both the TPM lattice and of the sublattice of additional
interactions the existence of finite hyperloops can be proved along
the following lines. Let us consider a lattice, TPM plus regular
sublattice, of finite size, with open boundary conditions. Think for
instance that our system is the portion of the lattice in
Fig.~\ref{fig:ITPM_lattice} which includes the $4$ additional
plaquettes there represented and the $36$ corresponding plaquettes of
the TPM, for a total number of $M=40$ plaquettes. With open boundary
conditions, the number of spins in this system is $N=49$.  In this
case nothing forbids the partition function to be exactly $\mathcal{Z}
= 2^N [\cosh(\beta)]^M$, since at zero temperature the corresponding
entropy is positive: $s(T=0)= 1-40/49 > 0$. In this case there is no
constraint implying the presence of hyperloops: it might be that for a
lattice of this size there is no finite hyperloop. If we think in
general to a lattice with the shape of a rhombus, with the same number
$L$ of TPM plaquettes on the horizontal and oblique side (consider
Fig.~\ref{fig:ITPM_lattice} to have an idea), and with open boundary
conditions, its partition function in absence of hyperloops is \be
\mathcal{Z} = 2^{(L+1)^2} [\cosh(\beta)]^{L^2
  +(L/3)^2}. \label{eq:ZnoLOOPS}\ee Due to the open boundary
conditions, if the number of plaquettes on each row is $L$ the number
of spins on the same row is $L+1$. On the other hand there are $L/3$
rows of additional plaquettes, each with $L/3$ plaquettes. This
explains why the total number of spins is $(L+1)^2$ and why the total
number of plaquettes is $L^2 +(L/3)^2$. For simplicity, let us
consider just the case when $L=3m$, where the integer number $m$
indicates therefore the number of plaquettes from the additional
sublattice in a row: the zero-temperature entropy per degree of
freedom corresponding to the free energy in Eq.~(\ref{eq:ZnoLOOPS}) is
\be s(T=0) = 1 - \frac{10 m^2}{(3m+1)^2}. \label{eq:SnoLOOPS} \ee 
From Eq.~(\ref{eq:SnoLOOPS}) we find the maximum value of the
parameter $m$ for which is possible to \emph{not} have hyperloops in
the partition function: $m^*=6$. For $m>m^*$ the expression of
$s(T=0)$ in equation Eq.~({\ref{eq:SnoLOOPS}) is negative: this
  actually proves (the system has discrete variables) that for
  \emph{any} lattice of size $m>m^*$ there must be hyperloops in the
  system. This way of reasoning allows one to prove not only that
  finite hyperloops are present, but also to fix an upper bound for
  the size of the smallest hyperloop. Since in a lattice with
  $m^*+1=7$ there is necessarily an hyperloop, then the smallest
  hyperloop cannot have more than $(m^*+1)^2=49$ plaquettes taken from
  the additional sublattice. Let us now explain why in this case there
  are non trivial contributions to the high-temperature expansion. In
  order to be general, we assume the smallest hyperloops is made of
  $k$ additional plaquettes of the sublattice and by $m$ plaquettes of
  the original TPM lattice. For every correlation function of $k'$
  plaquettes such that $k'$ is a multiple of $k$, and the plaquettes
  are all taken from the sublattice of additional ones, one must take
  into account disconnected hyperloops. The leading contribution of
  this kind of terms to the high-temperature expansion is
\bea 
\mathcal{Z} &=& 2^N[\cosh(\beta)]^{N+M_L} \bigg[ 1+ M_L [\thb]^{k+m} \nonumber \\ 
&+& M_L (M_L-k) [\thb]^{2k+2m} +  \nonumber \\ 
&+& M_L (M_L-k)  (M_L-2k) [\thb]^{3k+3m}\ldots \bigg], \nonumber \\
\label{eq:new-partition2}
\eea
where $M_L=\alpha_L N$. An explicit expression for the combinatorial
prefactor $\mathcal{C}(m,M)$ appearing in Eq.~(\ref{eq:ZGEN}) can be
provided for the terms appearing in Eq.~(\ref{eq:new-partition2}).
Let us indicate with
$\mathcal{C}(m_{\textrm{TPM}}+m_{\textrm{extra}},M)$ the combinatorial
factor which accounts for the multiplicity of an hyperloop with
$m_{\textrm{TPM}}$ plaquettes from the original TPM model and
$m_{\textrm{extra}}$ additional plaquettes. Then, according to the
expression in Eq.~(\ref{eq:new-partition2}) and taking $n$ a positive
integer we can write \be \mathcal{C}(k~n+m_{\textrm{extra}},M) \sim
M^n. \ee The series in Eq.(\ref{eq:new-partition2}) can be summed
leading to \be \mathcal{Z} ~\sim~ 2^N[\cosh(\beta)]^{N+M_L} e^{\alpha_L N [\thb]^{k+m}}, \ee 
\begin{figure}
\includegraphics[width=\columnwidth]{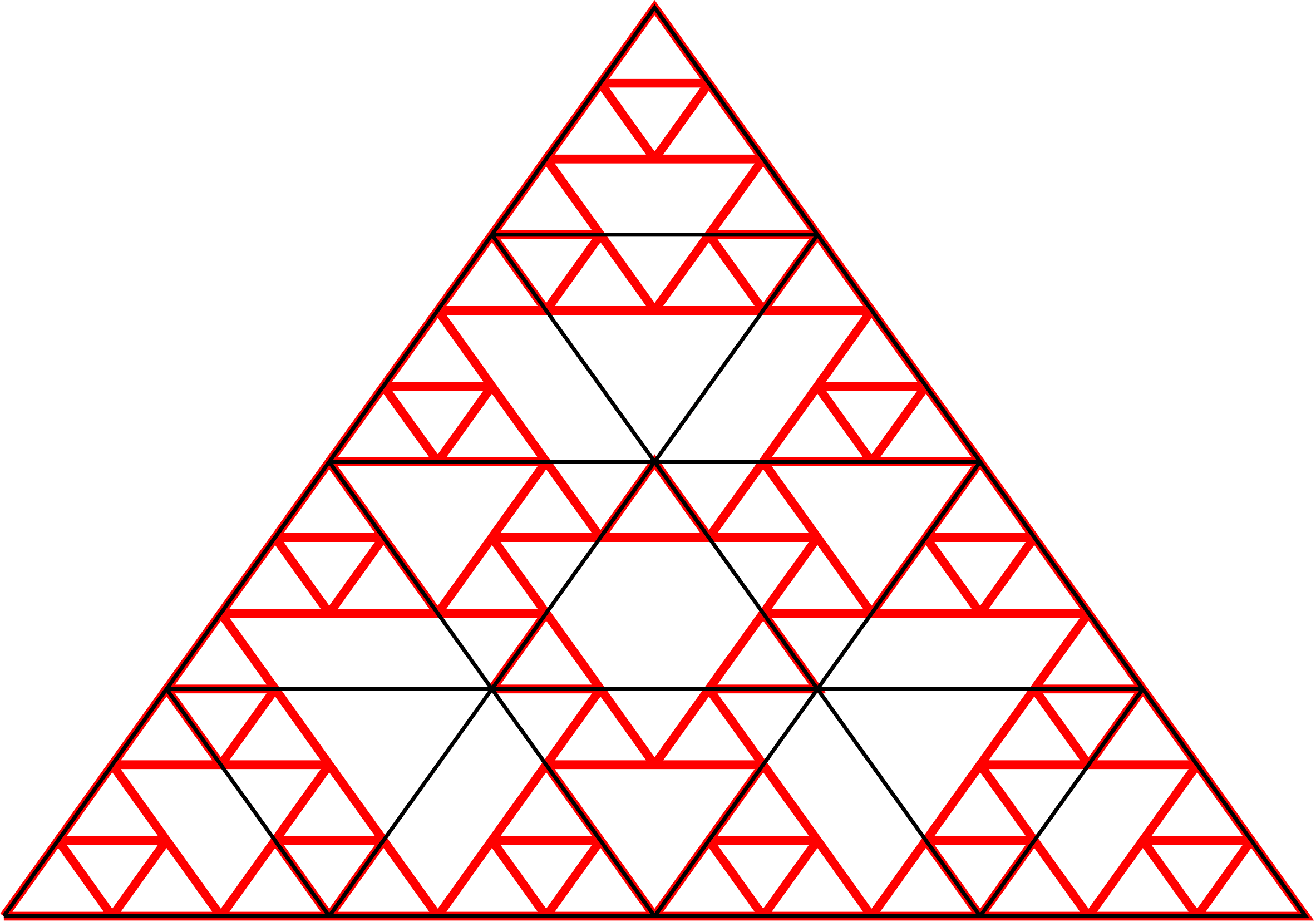}
\caption{(Color online) Smallest hyperloop of the high-temperature expansion
  for the modified TPM with extra short-range plaquettes discussed in
  Sec.~\ref{S1b}. In the picture can be counted 54 small plaquettes of
  the original TPM lattice, thick (red) triangles, and 10 plaquettes of
  the auxiliary sublattice, thin (black) triangles, which has as
  elementary cells equilateral triangles of side 3.}
\label{fig:smallest-diagram}
\end{figure}
which represents the contribution to the high-temperature expansion
provided by $k-$plaquettes connected hyperloops. We can conclude that
in the TPM the addition of short-range additional plaquettes always
introduce corrections to the high-temperature series: it is then hard
to say whether or not this expansion, which well reproduces the
properties of the system in the liquid phase, has positive entropy at
all temperatures. What numerical data (see below) show is that even
short-range additional interactions induce a first-order transition
with a rather robust supercooled liquid phase surviving below the
melting temperature. In order to characterize the hyperloop
corrections to the liquids phase, we looked to the smallest hyperloop
for the model defined at the beginning of this section. In order to do
this we used a simulated annealing method, explained in detail in
Appendix~\ref{A1a0}. According to simulated annealing, the smallest
hyperloop of the high-temperature series is the one shown in
Fig.~(\ref{fig:smallest-diagram}), which is made of $54$ plaquettes
from $H_{\textrm{TPM}}$ and $10$ plaquettes from
$H_{\textrm{extra}}$. Even if such hyperloop is ``big'', in the sense
that it provides a correction of the order $[\tanh(\beta)]^{64}$ to
the high-temperature expansion of the free-energy, it contains a
number of plaquettes from the additional sublattice ($10$ plaquettes)
well below the threshold of $(m^*+1)^2=49$ plaquettes (from the
additional sublattice) discussed above in this paragraph.\\ \\
The next question can then be: what about if we add to the TPM
Hamiltonian not all the plaquettes of the sublattice but just a
fraction of them randomly chosen? That is, what about if the sums in
Eq.(\ref{eq:new-partition}) runs over the plaquettes of a special
instance of the disorder, which corresponds to have in
$H_{\textrm{extra}}$ just a fraction $c$ of the plaquette of the
regular sublattice? The answer is that hyperloops will still be there,
just in a smaller amount compared to having \emph{all} the plaquettes
of the regular sublattice. Let us consider for instance the hyperloop
of Fig.~(\ref{fig:smallest-diagram}), made of $10$ close-by plaquettes
of the regular sublattice and $54$ plaquettes of the TPM. For a random
choice of the additional plaquettes such that a fraction $c$ of the
total number present in the sublattice is taken, in a high-temperature
expansion like the one in Eq.~(\ref{eq:new-partition2}) a contribution
as $ M_L [\tanh(\beta)]^{64}$ is replaced by $c M_L
[\tanh(\beta)]^{64}$, and correspondingly the correction to the
partition function becomes $\exp(c \alpha_L N [\tanh(\beta)]^{64})$.\\
The difference between adding randomly chosen long-range and
short-range plaquettes is at this point clear: while in the former
case it can be proved that the high-temperature expansion is trivial
and there is an entropic crisis, in the latter case there are
corrections coming from finite hyperloops, and these corrections are
most likely preventing any entropic crisis. For this case we present
the numerical evidence that, even with no clue on the presence of an
entropic crisis, the idea to add interactions to the TPM is a good
strategy to induce a non-trivial thermodynamics characterized by the
presence of a supercooled liquid phase.\\
\paragraph*{Numerical Simulations} In order to provide the numerical
evidence that even short-range interactions produce a non-trivial
thermodynamics we considered the case of a random choice of the
plaquettes on the regular sublattice and we found convenient also to
put weaker interactions on these additional plaquettes. Actually we
considered the Hamiltonian $H_{\textrm{extra}} = - J_1
\sum_{a=1}^{M_L} \sigma_{i_a} \sigma_{j_a} \sigma_{k_a}$, with
$J_1=1/5$ and with the sum running on a finite fraction of the
sublattice plaquettes. On one hand the random choice of a subset of
the plaquettes on the regular sublattice decreases the number of
finite hyperloops in the high-temperature expansion, on the other hand
a reduced strength of interactions on these plaquettes decreases the
weight of these hyperloops. Monte Carlo simulations done with $J_1=1$
showed that also with short-range plaquettes there is a melting
temperature $T_m$, but differently from the case of long-range
plaquettes (with a random choice of the spins) the system decays to
the ordered ground state as soon as $T<T_m$. Data are not reported
here, but we found that with the same annealing protocol used for
long-range plaquettes, namely temperature jumps of $\Delta T=0.04$
each $10^6$ or $10^7$ MC steps (depending on the temperature), the
system reaches the equilibrium ground states for all temperatures
$T<T_m$ with $M_L=0.1 N$ additional short-range plaquettes.  Before
comment the numerical data obtained with $J_1=1/5$, let us notice that
when the additional plaquettes are taken from a regular sublattice,
even for $T>T_m$ the high-temperature expansion yields a small but
finite magnetization. This can be clearly seen looking at the
high-temperature expansion of the magnetization:
\bea && \frac{1}{N}\left\langle
\sum_{i=1}^N \sigma_i \right\rangle \nonumber = \frac{1}{N\mathcal{Z}}
\sum_{i=1}^N \sum_{\boldsymbol \sigma}
\sigma_i~e^{-\beta(H_{\textrm{TPM}}[\boldsymbol
    \sigma]+H_{\textrm{extra}}[\boldsymbol \sigma])}= \nonumber \\ &=&
\frac{[\cosh(\beta)]^{M_L}}{N\mathcal{Z}} \left[ \sum_{i=1}^N
  \sum_{r=1}^{M_L} \langle \sigma_i~\sigma_{r_i} \sigma_{r_j}
  \sigma_{r_k}\rangle_{ _{\textrm{TPM}}} \tanh(J_1 \beta) + \ldots
  \right] \nonumber \\ &=& \frac{M_L}{N} \tanh^6(\beta)\tanh(J_1
\beta) + \ldots,
\label{eq:magn-high}
\eea 
\begin{figure}
\includegraphics[width=\columnwidth]{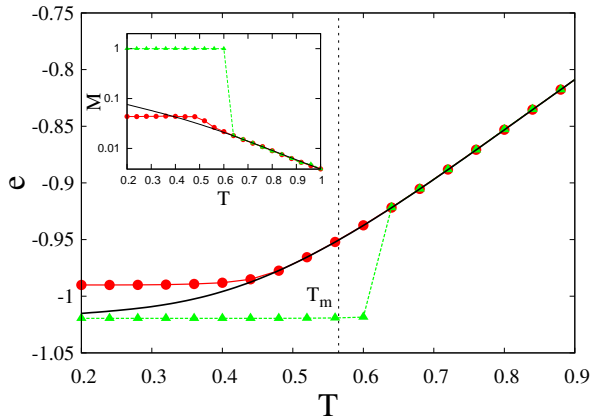}
\caption{(Color online) \emph{Main}: Energy hysteresis cycle for a TPM
  model with additional short-range plaquettes. We simulated the model
  on a lattice with the shape of a rhombus (see for instance the
  representation of Fig.~\ref{fig:ITPM_lattice}) with $L=128$
  plaquettes per side and a concentration of additional plaquettes
  $\alpha_L=0.1$: the coupling coefficient of the additional
  plaquettes is $J_1=1/5$. (Red) Circles: cooling (cooling rate from
  $10^6$ to $10^7$ MCS per $\Delta T=0.04$); (Green) Triangles:
  heating; Continuous (black) line: high-temperature expansion,
  $e(T)=-(\alpha_L \tanh(J_1/T)+\tanh(1/T))$. \emph{Inset}:
  Magnetization hysteresis cycle for the same model with the same
  cooling/heating protocol. (Red) Circles: cooling; (Green) Triangles:
  heating; continuous line: $m(T)$ from the high-temperature
  expansion: $m(T)=\alpha_L [\tanh(1/T)]^6\tanh(J_1/T)$, with
  $\alpha_L=0.1$ and $J_1=1/5$.}
\label{fig:hysteresis-SR}
\end{figure} 
which, as can be seen in Fig.~\ref{fig:hysteresis-SR}, nicely compares
with the results of numerical simulations. As is clear from
Eq.~(\ref{eq:magn-high}), the first effect of taking $J_1<1$ is to
reduce the high-temperature magnetization in the vicinity of $T_m$
making the supercooled liquid phase more stable. Let us briefly
explain how the equalities in Eq.~(\ref{eq:magn-high}) are
obtained. From the first to the second line of
Eq.~(\ref{eq:magn-high}) we just wrote down explicitly the
high-temperature expansion. The last identity of
Eq.~(\ref{eq:magn-high}) is clear if one recalls that each plaquette
of $H_{\textrm{extra}}$ is represented by a triplets of spins
$\sigma_{r_i} \sigma_{r_j} \sigma_{r_k}$ at the vertices of an
equilateral triangle of side $3$. It can be easily realized that those
spins form an \emph{hyperfield}~\cite{RWZ01} when multiplied with the
plaquettes of $H_{\textrm{TPM}}$ which are enclosed within the
perimeter of the same triangle.  Borrowing the terminology
of~\cite{RWZ01} we call \emph{hyperfield} a set of interactions such
that all spins but one, $\sigma_i$, appear an even number of
times. Since the plaquettes of $H_{\textrm{extra}}$ are all
represented by triplets of spins placed at the corners of equilater
triangles of side $3$, each of them contributes an hyperfield
$\sigma_i$ in the high-temperature expansion, and this hyperfield
becomes an hyperloop of weight $[\tanh(\beta)]^6\tanh(J_1 \beta)$ when
multiplied for the corresponding spin $\sigma_i$ appearing in the
definition of the magnetization. Then, by taking also the
approximation $\mathcal{Z}\sim 2^N [\cosh(\beta)]^{N+M_L}$ (which is
exact in the case of long-range plaquettes, see Sec.~\ref{S1a}), we
have the result of the last line in Eq.~(\ref{eq:magn-high}). \\ Our
numerical study showed that for the smaller value of the coupling
constant $J_1=1/5$, a number $M_L= 0.1 N$ of additional plaquettes and
the same annealing protocol already mentioned above in this section,
at all the temperatures studied the system does not decay to the
ordered ground state, see Fig.~\ref{fig:hysteresis-SR}. In the inset
of Fig.~\ref{fig:hysteresis-SR} is shown the behaviour of the
magnetization, which in the paramagnetic phase is well reproduced by
the high-temperature expansion. The model with randomly chosen
plaquettes on a regular sublattice seems therefore a good one to
reproduce the standard scenario of realistic glass-formers: a
first-order phase transition to an ordered ground state plus a
long-lived supercooled liquid phase. A system where a similar
behaviour is found is the Coupled Two Level System (CTLS) model~\cite{CGG03a,CGG03}. The CTLS is
a non-disordered plaquette model which presents, as our TPM with
additional plaquettes, a first-order transition to an ordered ground
state and a metastable supercooled liquid phase. To summarize, we can
say that even when is not possible to argue about the existence of a
thermodynamic glass transition, the addition of short-range plaquettes
to the TPM induce a first-order phase transition and the formation of
a robust supercooled liquid phase. We can therefore argue that even
with short-range interactions the presence of a non trivial
thermodynamics is controlled by the ratio $\alpha=M/N$ between the
number of plaquettes and the number of spins. In plaquette models with
short-range interactions the glass-forming ability is then a matter of
competition between two time scales: the time scale to nucleate the
the crystal and the relaxation time of the supercooled liquid. For a
detailed discussion on how a stable supercooled liquid phase can be
obtained by appropriately tuning the cooling rate procedure let us
refer the reader to~\cite{CGG03a,CGG03,AC09}.
\section{Random-Diluted TPM: phase diagram at $T=0$}
\label{S2}


The analysis of the present section is based on the deep connection
between the thermodynamics of plaquette models and the properties of
solutions of the XOR-SAT problem. The latter is a constraint
satisfaction problem which has been very successfully described within
the landscape scenario of the ideal glass transition~\cite{mrz03}. The
relation between the TPM and the XOR-SAT comes from the fact that the
ground states of a TPM can be obtained as the solutions of a
XOR-SAT. By assuming the change of variables $\sigma_i=(-1)^{n_i} $, a
ground state of the TPM can be always written as the solution of the
system of linear equations: \be \left( \sum_{i\in \partial p} n_{i}
\right)_{\textrm{mod} 2} = 0, ~~~~~~~ \forall ~p \label{eq:XOR} \ee
where with the notation $i\in \partial p$ we indicate the spins in the
plaquette $p$. The ground states of the TPM are represented by
\emph{all} the solutions of the system in Eq.~(\ref{eq:XOR}), which is
made of $M$ linear equations, the plaquettes, in $N$ variables, the
spins. The topology of the interaction network is specified by the
equations of the linear system. For instance, in the triangular 2D
lattice of the TPM each spin $n_i$ appears in three equations and in
each equation it is coupled to the two other spins belonging to the
same plaquette. When the network formed by spins and plaquettes
corresponds to a random hyper-graph (which is a locally tree-like
network with loops of order $\log(N)$) it is known that the properties
of the solutions of the XOR-SAT are fully determined by the parameter
$\alpha=M/N$~\cite{mrz03}. For a TPM on the random graph, even with
non-disordered interactions~\cite{fmrwz01}, two transitions are found
varying $\alpha$, respectively at $\alpha_d$ and
$\alpha_{\textrm{unsat}}$. When $\alpha<\alpha_d$ the system of
equations in Eq.~(\ref{eq:XOR}) is solvable and happens that few spins
flips are sufficient to go from one solution to another: the set of
solutions forms a unique cluster. On the contrary, when $ \alpha_d <
\alpha < \alpha_{\textrm{unsat}}$ the set of solutions splits in
clusters separated by extensive barriers: it is necessary to flip an
extensive number of spins to pass from one cluster to the
other. Finally, when $\alpha>\alpha_{\textrm{unsat}}$ the system in
Eq.~(\ref{eq:XOR}) is no more solvable: at $\alpha_{\textrm{unsat}}$
takes place the SAT/UNSAT transition. With respect to any random
update algorithm designed to move across the phase space of variables
$\lbrace n_i \rbrace_{i=1,\ldots,N}$, the SAT/UNSAT transition
represents an ideal glass transition. The non-satisfiable phase of the
XOR-SAT coincide with the glass phase of the related TPM: as a
consequence, when $\alpha > \alpha_{\textrm{unsat}}$ the TPM (on
random graph) is in the glass phase at $T=0$, which means that it has
an ideal glass transition at $T>0$.

\subsection{Leaf-removal algorithm and $T=0$ phase diagram}
The ``leaf-removal'' algorithm is a decimation scheme used to study
the satisfiability of the XOR-SAT model on random regular
graphs~\cite{mrz03}. We discuss here how this algorithm can be used to
investigate the properties of our Random-Diluted TPM. It is called
``leaf'' every spin which appears in only one plaquette (i.e. every
variable appearing in a single equation of the XOR-SAT problem):
``leaf-removal'' is a prescription to remove iteratively from the
graph all the spins which are (or become) leaves. The procedure is
iterative because after the removal of each leaf new leaves may appear
in the system. When $\alpha < \alpha_d$ leaf-removal is able to remove
all the spins from the graph, while for $\alpha>\alpha_d$ the
algorithm stops leaving the so called ``core'', i.e. a set of spins
among which no one is a leaf. The clustering of solutions of the
XOR-SAT at $\alpha_d$ corresponds to the formation of the core. The
SAT/UNSAT transition takes place when the number of equations
(plaquettes) in the core, $M_c$, becomes larger than the number of
variables left on it, $N_c$.  The critical value
$\alpha_{\textrm{unsat}}$ can be determined by studying the ratio
$M_c/N_c=\alpha_c$ in the core, and corresponds to $\alpha_c=1$. On a
random regular graph the dependence of $M_c$ and $N_c$ on $\alpha$ can
be determined analytically in the thermodynamic limit~\cite{mrz03}.
The leaf-removal algorithm can be used to study the formation of the
core and the behaviour of $\alpha_c$ on the core for a XOR-SAT on
every kind of topology. Nevertheless, only for random graphs it is
proven that the formation of the core corresponds to the clustering
transition and the value $\alpha_c=1$ to the SAT/UNSAT
transition~\cite{mrz03}.  On finite dimensional topologies a proof of
this correspondence is still lacking. This notwithstanding, we studied
numerically the action of the leaf-removal on our Random-Diluted TPM,
proposing a ``tentative'' phase diagram. We compare this numerical
phase diagram with that obtained by analytically solving leaf-removal
for the representation of our Random-Diluted TPM on the random graph:
in this case the leaf-removal analysis yields exactly the
thermodynamic properties of the system~\cite{mrz03}. Let us stress
that by running the leaf-removal algorithm in a finite-dimensional
geometry, we may find both a critical value $\alpha^*$ for the
formation of the core and a critical value $\alpha^{**}$ where
$M_c/N_c=1$, but there is not proof that they correspond respectively
to the clustering and UNSAT transition. This is the reason why the
phase diagram of Fig.~\ref{fig1} is just tentative. The conjecture
that even in finite dimensions at $\alpha^{**}$ really takes place the
UNSAT transition, namely $\alpha^{**}=\alpha_{\textrm{unsat}}$, is
supported within our analysis just by the agreement we find between
the numerical finite-dimensional and exact mean-field predictions on
the phase diagram.

Until now we have discussed about a single parameter $\alpha$, but for
the Random-Diluted TPM and its representation on the random graph we
need two: $\alpha_s$, which represents the concentrations of
plaquettes in the 2D lattice, and $\alpha_L$, which represents the
concentration of long-range plaquettes. The Random-Diluted TPM
corresponds in practice to a random-graph structure, parametrized by
$\alpha_L$, built on the top of a diluted two-dimensional network,
characterized by a dilution parameter $\alpha_s$. The probability that
a spin is attached to $\ell$ short-range plaquettes depends on
$\alpha_s$ and is: \be \rho_\ell(\alpha_s) =
\binom{3}{\ell}\alpha_s^\ell(1-\alpha_s)^{3-\ell}, \label{eq:dist1}
\ee while the probability that a spin is attached to $\ell$ long-range
plaquettes depends on $\alpha_L$ and reads: \be p_{\ell}(\alpha_L) =
e^{-3\alpha_L}\frac {(3\alpha_L)^{\ell}}{\ell!}. \label{eq:dist2} \ee
The probability that a spin is attached to overall $\ell$ plaquettes
is \be n_\ell= \mathcal{N}^{-1} \sum_{r=0}^{\ell_0} \rho_r(\alpha_s)
p_{\ell-r}(\alpha_L), \label{eq3} \ee with \bea
  \mathcal{N} &=& \sum_{\ell=0}^\infty \sum_{r=0}^{\ell_0}
  \rho_r(\alpha_s) p_{\ell-r}(\alpha_L) = \nonumber \\ &=&
  1-\sum_{r=0}^3 \sum_{k=r}^3 \rho_r(\alpha_s)
  p_{k-r}(\alpha_L)+\sum_{k=0}^3\sum_{r=0}^k \rho_r(\alpha_s)
  p_{k-r}(\alpha_L), \nonumber \\ \eea where
$\ell_0=\textrm{min}\lbrace \ell,3\rbrace$. When the Random-Diluted
TPM is represented on the random graph the small-word topology,
i.e. the two-dimensional lattice plus few long-range connections, is
lost. The only ingredient of the original model which is kept is the
presence of two kinds of plaquettes, each characterized by a different
probability for the connectivity with the spins. For the
representation on the random graph the adjectives ``short-range'' and
``long-range'' are therefore only conventional: the former denotes the
plaquettes attached to spins with the probability of
Eq.~(\ref{eq:dist1}), the latter plaquettes attached to spins with the
probability of Eq.~(\ref{eq:dist2}).\\
\begin{figure}
\includegraphics[width=\columnwidth]{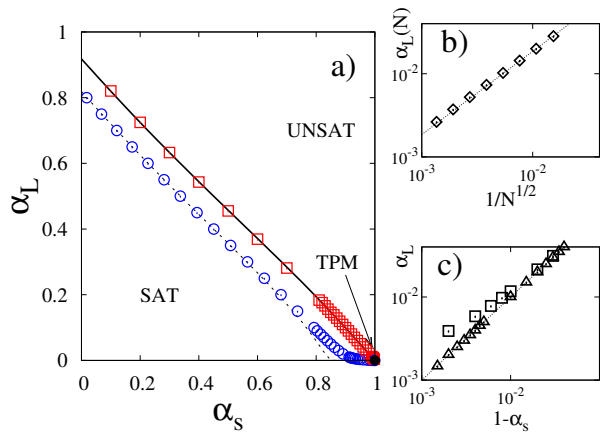}
\caption{(Color online) \underline{Panel \emph{a)}}: ``$T=0$'' phase
  diagram of the Random-Diluted TPM (see Eq.~(\ref{eq3})). Symbols
  represent numerical data obtained running the leaf-removal algorithm
  on finite dimensional geometries, lines the results of analytical
  calculation on the Bethe lattice. Circles (blue) and the dotted line
  represents the critical line for the formation of the core, squares
  (red) and the continuous line are the critical line for the
  SAT/UNSAT transition.  \underline{Panel \emph{b)}}: Diamonds,
  numerical estimate in finite dimensions of the critical value
  $\alpha_L$ for the SAT/UNSAT transition as function of $\sqrt{N}$ ,
  with $N$ the size of the system, at fixed $\alpha_s=1$; dotted line,
  linear fit of data. \underline{Panel \emph{c)}}: zoom of the
  SAT-UNSAT transition line close to the point
  $(\alpha_s=1,\alpha_L=0)$.  Triangles, analytic result on the random
  graph; Squares, numerical data for a system with $N=10^6$ spins;
  Dotted line, linear fit of analytic prediction} \label{fig1}
\end{figure}
Fig.~\ref{fig1} shows the phase diagram of the Random-Diluted TPM
obtained by running the leaf-removal algorithm in finite dimensions
and that obtained by solving the corresponding equations (see
Appendix~\ref{A2}) in the mean-field (random graph) approximation.  In
both cases we determine a ``critical'' line
$(\alpha_s^{\textrm{core}},\alpha_L^{\textrm{core}})$ for the
formation of the core (clustering transition on the random graph) and
a critical line
$(\alpha_s^{\textrm{unsat}},\alpha_L^{\textrm{unsat}})$ from the
condition $M_c/N_c=1$ (unsat transition on random graph). First of
all, let us note the agreement between numerical and analytical
predictions on the location of the line
$(\alpha_s^{\textrm{unsat}},\alpha_L^{\textrm{unsat}})$ in
Fig.~\ref{fig1}. Since with $\alpha_s=0$ the Random Diluted-TPM is
perfectly equivalent to the XOR-SAT~\cite{mrz03}, for $\alpha_s=0$ we
recover the random graph result
$\alpha_L^{\textrm{unsat}}=0.918$~\cite{mrz03}. By looking at the left
part of the phase diagram in Fig.~\ref{fig1} we are indeed not
surprised that for $\alpha_L\sim 1$ and $\alpha_s\ll 1$ the analytic
predictions on the random graph are in agreement with the numerical
analysis in finite dimensions: the Random-Diluted TPM is almost a
random graph for these values of the parameters. What is more
surprising is to find an agreement between the numerical and the
analytical estimate of
$(\alpha_s^{\textrm{unsat}},\alpha_L^{\textrm{unsat}})$ in the bottom
right part of the phase diagram in Fig.~\ref{fig1}, where $\alpha_L\ll
1$ and $\alpha_s\sim 1$. In this region the Random-Diluted TPM is
almost a two-dimensional model, while the analytical predictions are
given for a random graph geometry. It really looks like that as soon
as few long-range interactions are added to the TPM its physics
becomes immediately well represented by the random-graph. If the true
thermodynamics of the Random-Diluted TPM could be predicted on the
basis of the analytic results of Fig.~\ref{fig1}, we would say that as
soon as \emph{any} finite concentration $\alpha_L$ of random
plaquettes is added to the TPM the model develops a finite-temperature
glass transition. We do not still have a proof of that, therefore by
now this is just a conjecture supported by the agreement between
numerics and analytics on the location of the
$(\alpha_s^{\textrm{unsat}},\alpha_L^{\textrm{unsat}})$ line. The
small panels of Fig.~\ref{fig1} illustrate the finite-size scaling
analysis needed to assess the agreement of analytical and numerical
predictions on the behaviour of the
$(\alpha_s^{\textrm{unsat}},\alpha_L^{\textrm{unsat}})$ transition
line close to the point $(\alpha_s=1,\alpha_L=0)$.\\ Concerning the
analytical and numerical data on the location of the
$(\alpha_s^{\textrm{core}},\alpha_L^{\textrm{core}})$ line in the
bottom right part of the phase diagram of Fig.~\ref{fig1} we find a
certain disagreement: two main comments on this are in order. First,
it is well known that the clustering transition which takes place at
$\alpha_d$ on the random graph, and which corresponds to dynamical
ergodicity breaking, is a purely mean-field phenomenon which turns
into a crossover in finite dimensions. From this point of view we are
not concerned about the disagreement found between the analytical
(mean-field) and numerical (finite-dimensional) results on the
position of the line
$(\alpha_s^{\textrm{core}},\alpha_L^{\textrm{core}})$. On the other
hand, the numerical results on the location of the line
$(\alpha_s^{\textrm{core}},\alpha_L^{\textrm{core}})$ are \emph{per
  se} interesting and represent a useful source of information on the
model, as will be discussed in the next section Sec~\ref{S2b}.\\
\begin{figure}[t!]
\includegraphics[width=\columnwidth]{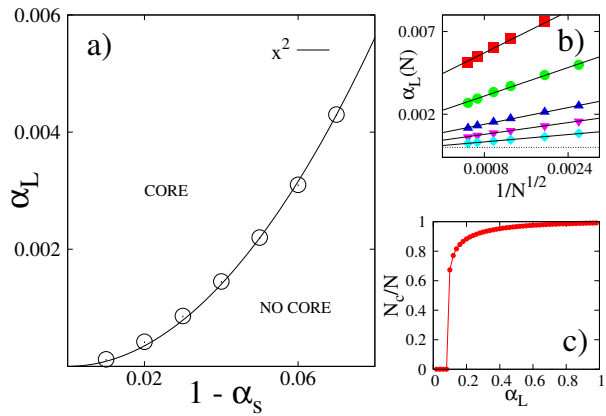}
\caption{(Color online) \underline{Panel~\emph{a)}}: Circles are the
    critical values $(\alpha_s,\alpha_L)$ for the formation of the
    core extrapolated in the thermodynamic limit from numerical data
    at finite $N$. The continuous line represent a quadratic fit of
    data $y = a x^2$, with prefactor $a=0.877$, see also
    Eq.~(\ref{eq:parametric}) in
    Sec.~\ref{S2b}. \underline{Panel~\emph{b)}}: critical values
      $\alpha_L(N)$ for core formation as unction of the size $N$ of
      the system for different values of $\alpha_s$ (different
      symbols), from top to bottom: $\alpha_s=$0.93, 0.95, 0.97, 0.98,
      0.99. \underline{Panel~\emph{c)}}: concentration of spins in the
        core as function of $\alpha_L$ for a fixed value of
        $\alpha_s=0.8$, with $N=1024^2$.}
\label{fig:dyn-scaling}
\end{figure}
\subsection{Leaf-removal and exact calculation of $\mathcal{Z}$} 
\label{S2b}

The numerical estimate of the position of the line
$(\alpha_s^{\textrm{core}},\alpha_L^{\textrm{core}})$ in the phase
diagram of Fig.~\ref{fig1} is quite interesting even in the case of a
finite dimensional geometry. In this case the line
$(\alpha_s^{\textrm{core}},\alpha_L^{\textrm{core}})$ is the upper
boundary of the region where the partition function of the
Random-Diluted TPM can be exactly calculated and has the expression
\begin{equation} 
\mathcal{Z} = 2^N [\cosh(\beta)]^{M_s+M_L} = \left( 2~[\cosh(\beta)]^{\alpha_s+\alpha_L} \right)^N. 
\label{eq:free-partition}
\end{equation}
The exact explanation of why leaf-removal allows us to check whether
or not the partition function can be exactly summed, yielding the
expression in Eq.~(\ref{eq:free-partition}), can be found in
Appendix~\ref{A2b}. The behaviour of the
$(\alpha_s^{\textrm{core}},\alpha_L^{\textrm{core}})$ line (numerical
data) close to the point $(\alpha_s=1,\alpha_L=0)$ in the phase
diagram of Fig.~\ref{fig1}, which correspond also to the data in the
main panel of Fig.~\ref{fig:dyn-scaling}, shows that, among all the
two-dimensional models belonging to the line $(\alpha_s,0)$, the only
one such that \emph{any} concentration $\alpha_L>0$ of long-range
plaquettes is ``critical'' is the original TPM. As ``critical'' we
mean that the possibility to exactly compute $\mathcal{Z}$ according
to the expression in Eq.~(\ref{eq:free-partition}) is spoiled as soon
as any finite concentration of long-range plaquettes, $\alpha_L>0$, is
introduced. In order to make ourselves really sure about that we need
to know the behaviour of the system in the thermodynamic limit. In the
top right panel of Fig.~\ref{fig:dyn-scaling} is presented the study
of finite size effects, while in the main panel of
Fig.~\ref{fig:dyn-scaling} appears the resulting estimate for some
points of the line
$(\alpha_s^{\textrm{core}},\alpha_L^{\textrm{core}})$. These point can
be well interpolated with a parabola:
\begin{equation}
\alpha_L^{\textrm{core}} + \mO(N^{-1/2})\sim
(1-\alpha_s^{\textrm{core}})^2.
\label{eq:parametric}
\end{equation}
The parabolic fit of the data in the main panel of
Fig.~\ref{fig:dyn-scaling} represents the main evidence that the only
model on the line $(\alpha_s,\alpha_L=0)$ such that the addition of
extra plaquettes is critical is the TPM model. To conclude this
section let us notice that also in an almost finite-dimensional
geometry, $\alpha_s=0.8$ and $\alpha_L\sim 1$, we find that the
formation of the core is a discontinuous process, see panel c) of
Fig.~\ref{fig:dyn-scaling}, as is usually found for random
geometries. This result supports the view that the physical properties
of the Random-Diluted TPM can be well represented even on a random
graph. Our finding that the formation of the core happens
discontinuously upon changing $\alpha_L$ also in finite dimensions is
intriguing: it is not the first time that a similar discontinuous
transition has been observed in finite
dimensions~\cite{TBF06,TB08,TB08b,JS10,GTS14}. The spiral model
of~\cite{TB08,TB08b}, which is a KCM, is the example of a
finite-dimensional system where a dynamical transition really takes
place and is due to the formation in the system of an infinite compact
cluster of ``frozen'' (i.e. not allowed to move due to the kinetic
constraint) spins. A cluster of spins which cannot flip due to a
kinetic constraint is not exactly reconducible to the leaf-removal
``core'' of our Random-Diluted TPM. Yet, if the mean-field scenario is
predictive also for the behaviour in finite-dimensions, the formation
of the core should take place when, in order to move in phase-space,
we need to flip an extensive amount of spins, which we may very
roughly think about as a ``frozen cluster''.\\

\section{Random-Diluted TPM on random graph: phase diagram at $T>0$.}
\label{S3}

In this section we present results on the finite-temperature phase
diagram of the Random Diluted-TPM model on the random regular graph
(Bethe lattice), where the temperature $T_K$ of the ideal glass
transition can be computed exactly.

According to the presence of both long and short-range plaquettes in
the Random-Diluted TPM, the cavity equations (Appendix~\ref{A3}) for
its representation on the random graph are written by means of two
different cavity fields, as is usually done for small-word
networks~\cite{wnh05}. The field $u_{\alpha\rightarrow i}$ determines
the probability distribution $p(\sigma_i)\sim e^{u_{\alpha\rightarrow
    i}\sigma_i}$ when all the plaquettes attached to $\sigma_i$ but
$\alpha$ are removed, and $\alpha$ is a long-range plaquette. The
field $v_{\alpha\rightarrow i}$ determines the probability
distribution $p(\sigma_i)\sim e^{v_{\alpha\rightarrow i}\sigma_i}$
when all the plaquettes attached to $\sigma_i$ but $\alpha$ are
removed, and $\alpha$ is a short-range plaquette. The cavity
equations, written and discussed in Appendix~\ref{A3}, allow us to
find the equilibrium values of the fields $u$ and $v$, from which all
the thermodynamic potentials can be calculated (formulas are in
Appendix~\ref{A3}). For fixed values of $\alpha_s$ and $\alpha_L$, the
glass transition temperature $T_K$ is obtained as the temperature
where the configurational entropy $\Sigma$ vanishes.  
\begin{figure}
\includegraphics[width=\columnwidth]{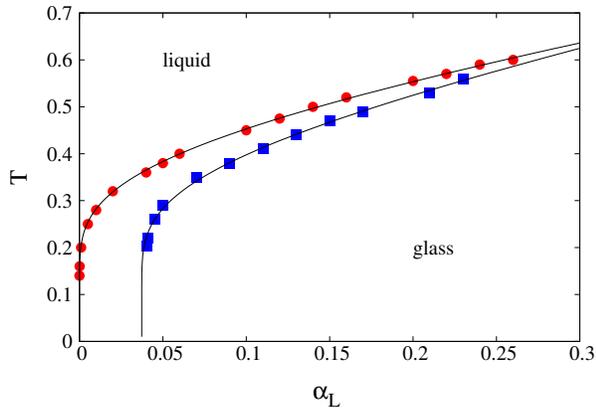}
\caption{(Color online) Phase diagram of the Random Diluted-TPM model
  on the Bethe lattice in the $(\alpha_L,T)$ plane for two different
  values of dilution $\alpha_s$: Circles (red), $\alpha_s=1$; Squares
  (blue), $\alpha_s=0.96$; Continuous lines: fits of the data with the
  function $\alpha_L(T)= C_1 \exp(-C_2/T) + \alpha_L(0)$, where $C_1$
  and $C_2$ are fit parameters.}
\label{fig2}
\end{figure}
In Fig.~\ref{fig2} is represented the phase diagram of the
Random-Diluted TPM in the plane $(\alpha_L,T)$ for two values of the
short-range plaquettes concentration: $\alpha_s=1$ and
$\alpha_s=0.96$. While for $\alpha_s=1$ the glass transition
temperature vanishes when also the concentration $\alpha_L$ of
long-range plaquettes vanishes, when $\alpha_s=0.96$ we find that
$T_K$ vanishes at a finite value of the additional plaquettes
concentration, $\alpha_L^\textrm{min}>0$: when $\alpha_L$ belongs to
the interval $[0,\alpha_L^\textrm{min}]$ the system is liquid at all
the temperatures. By definition, the finite-temperature phase diagram
of the Random-Diluted TPM on the Bethe lattice must agree with the
analytical solution of the leaf-removal algorithm, that is also
obtained on the random graph. From the leaf-removal analysis of
Sec.~\ref{S2} we already know that for all the concentrations of
short-range plaquettes $\alpha_s<1$ there is always a value
$\alpha_L^\textrm{min}>0$ such that for concentrations of long-range
plaquettes $\alpha_L<\alpha_L^\textrm{min}$ the system is liquid at
all temperatures. This happens because, as we discussed in
Sec.~\ref{S2} (see also Fig.~\ref{fig1}), the only value of $\alpha_s$
such that the related XOR-SAT problem becomes UNSAT (glass phase) for
every $\alpha_L>0$ is just $\alpha_s=1$. Let us note that the phase
diagram of our model in the plane $(\alpha_L,T)$, i.e. for a fixed
concentration $\alpha_s$ of short-range plaquettes (see
Fig.~\ref{fig2}), has a remarkable similarity with the phase diagram
in the plane $(\epsilon,T)$ of~\cite{G14}, where $\epsilon$ is an
external field coupling different replicas of a TPM. According
to~\cite{G14} the critical line $T_K(\epsilon)$ approaches the origin
with infinite slope~\cite{G14}, $T_K \sim [\log(\epsilon)]^{-1}$: we
are going to show that the same happens in our case to
$T_K(\alpha_L)$, when the concentration $\alpha_L$ of long-range
plaquettes is sent to zero. We want to emphasize that $\alpha_L$ plays
a role analogous to $\epsilon$. The argument for the behaviour of
$T_K(\alpha_L)$ in our system is rather simple and is exact on the
random graph. From~\cite{mrz03} we know that in the UNSAT phase the
ground states of a TPM on random graph have extensive energy. The
energy of the system is by definition the sum over plaquette energies,
so that when a ground state has extensive energy it means that there
is an extensive amount of excited plaquettes in it. To have ground
states with extensive energy is equivalent to have a minimum value,
$\varepsilon^{\textrm{min}}$, for the energy per plaquette. In the TPM
the concentration of defects at low temperatures behaves as $c \sim
e^{-2\beta}$ and we can assume that in presence of a small amount of
extra plaquettes this dependence is roughly the same, say $ c \sim
e^{-a 2 \beta}$ with $a \sim 1$. Since the existence of a minimum
value $\varepsilon^{\textrm{min}}$ is equivalent to a minimum
concentration of excited plaquettes, also a minimum $ c^\textrm{min}$
is fixed. From the constraint of a minimum allowed concentration of
excited plaquettes we can define a critical temperature $T_K$ as
$c^\textrm{min} = e^{-a 2 /T_K}$.  Clearly the minimum value of the
energy per plaquette and the minimum concentration of excited
plaquettes must be proportional $c^\textrm{min} \sim
\varepsilon^\textrm{min}$. By looking at the behaviour of the energy
in the ground states as a function of $\alpha$ presented
in~\cite{mrz03}, it is reasonable to assume also for our
Random-Diluted TPM on random graph that we have
$\varepsilon^\textrm{min}\sim \alpha_L^b$, with $b\sim 1$. Putting
together all the information collected above we can conclude that
$\alpha_L^b = e^{-a 2 /T_K}$, which in turn implies
\be T_K = -\frac{a}{b}\frac{2}{\log(\alpha_L)}. \label{eq:logarithmic} \ee
Eq.~(\ref{eq:logarithmic}) allows us a good fit of the data in
Fig.~\ref{fig2} and accounts for the infinite slope of the curve
$T_K(\alpha_L)$ approaching $\alpha_L=0$. This infinite slope is
signaling that around $\alpha_L=0$ the Random-Diluted TPM is
sensitive to arbitrarily small perturbations, which induce the
formation of glass-phase. The same happens for the original
two-dimensional Triangular Plaquette Model in presence of an external
field $\epsilon$ coupling replicas~\cite{G14}.

\section{Conclusions}
\label{S4}

In this paper we studied the thermodynamic properties of the
Triangular Plaquette Model in presence of additional plaquettes,
namely we looked to what happens when $\alpha=M/N>1$, where $M$ is the
number of plaquettes and $N$ the number of spins.  We have
demonstrated that in the small-word lattice obtained by adding
long-range plaquettes to the TPM the high-temperature expansion of the
free-energy, which can be computed in the annealed approximation, has
an entropic crisis. In the same model we find the numerical evidence
of a first-order transition to a an ordered phase at $T_m$, with a
remarkably stable supercooled liquid phase at lower temperatures. The
same phenomenology is found also when the additional interactions are
short-range, although in this case there are corrections to the
high-temperature expansion which very likely prevent the entropy
crisis. Since the presence of an ideal glass transition is more likely
when the entropy crisis takes place, we studied in more detail in the
rest of the paper the model with long-range additional plaquettes,
i.e. the Random-Diluted TPM. Our deepest investigation of the
thermodynamic properties of the Random-Diluted TPM was carried on by
means of the leaf-removal algorithm, usually applied to constraint
satisfaction problems~\cite{mrz03}. The advantage of leaf-removal is
that it allows to infer the thermodynamic properties of the model just
analyzing the interaction network. The drawback is that the
correspondence between the properties of the interaction network and
the thermodynamics is exact only for random geometries. By means of
the leaf-removal algorithm we obtained a \emph{tentative} phase
diagram in the space of parameters $(\alpha_s,\alpha_L)$, where
$\alpha_s$ is the concentration of plaquettes in the 2D triangular
lattice while $\alpha_L$ is the concentration of ``long-range''
plaquettes. Such a phase diagram suggests that among the 2D plaquette
models with different dilutions $\alpha_s$ the original
TPM~\cite{nm99} ($\alpha_s=1$) is the only one where the addition of
\emph{any} concentration of long-range plaquettes makes the
thermodynamic non-trivial. That is why we say that the TPM is
stochastically \emph{unstable}: arbitrarily small perturbations of the
Hamiltonian have dramatic effects on the thermodynamic properties of
the model. Moreover, our results suggest that even in finite
dimensions the parameter that controls this \emph{stochastic
  unstability} is the ratio $\alpha$ between the number of plaquettes
and the number of spins. These considerations are also compatible with
the results of~\cite{G14,TJG15}. In~\cite{G14,TJG15} is shown how the
TPM supports both dynamic and thermodynamic phase transition under the
influence of arbitrarily small external fields. We find remarkable the
similarity between the behaviour of the glass transition temperature
$T_K$ as function of $\alpha_L$ in our Random-Diluted TPM (on the
random graph) and as function of $\epsilon$ in~\cite{G14}: in both
situations an infinitesimal \emph{amount of perturbation} lifts the
critical temperature $T_K$ to finite values. We can conclude saying
that further investigations on the stochastic stability of the TPM are
mandatory.
\begin{acknowledgments}
We thank J.-P. Garrahan, F. Ricci-Tersenghi and R. Santachiara
for useful discussions. GG acknowledges support from the project NSTGEA
of Labex PALM.
\end{acknowledgments}

\appendix 

\section{High-temperature expansion with additional plaquettes}
\label{A1}

\subsection{Random choice of spins in the new plaquettes: Random-Diluted TPM} 
\label{A1a}

\subsubsection{{\bf Triviality of the high-temperature expansion}}
\label{A1aa}

In order to say that the high-temperature expansion of
Eq.~\ref{eq:high-temp} in Sec.~\ref{S1a} is trivial we need to show
that \be \lim_{N\rightarrow\infty} \binom{M_L}{m} \langle
\tau_1\ldots\tau_m \rangle_{ _{\textrm{TPM}}} = 0 ~~~~~~~\forall~m
\label{eq:TermeHighTSerie}
\ee Let us recall that the correlation function in
Eq.~(\ref{eq:TermeHighTSerie}) is the multispin correlation function
$\langle \sigma_1 \ldots \sigma_{3m} \rangle_{ _{\textrm{TPM}}}$,
where the $3m$ spins are randomly chosen with uniform probability on
the 2D lattice.  The multispin correlation function $\langle \sigma_1
\ldots \sigma_{3m} \rangle_{ _{\textrm{TPM}}}$ is different from zero
only when is possible to find plaquettes in the TPM such that a set of
one or more hyperloops can be formed which include all the $3m$
spins. Since in the TPM we know that both the magnetization $\langle
\sigma_i \rangle_{ _{\textrm{TPM}}}=0$ and the two spin correlation
function $\langle \sigma_i \sigma_j \rangle_{ _{\textrm{TPM}}}=0$ are
zero, each of the hyperloops contains necessarily at least three
spins. At the same time, since the $3m$ spins are chosen with uniform
probability on the lattice, the typical distance between any two of
them is $\mathcal{O}(\sqrt{N})$. This in turn means that the typical
the distance between any two spins which belong to the same hyperloop
is also $\mathcal{O}(\sqrt{N})$. Then, due to the fact that we
consider the thermodynamic limit at fixed $m$, any hyperloop which
contains the $3m$ spins of the the additional plaquettes contains also
an infinite number of plaquettes of the TPM model. More precisely,
since the hyperloops connects spins at a distance
$\mathcal{O}(\sqrt{N})$, the number of plaquettes of the TPM in the
hyperloop is $\mathcal{O}(N^{d_{\textrm{H}}/2})$, where $d_H$ is the
fractal dimension of the Sierpinski gasket in D=2. One then has that
the value of the correlation function between $3m$ spins chosen with
uniform probability on the lattice is dominated at large $N$ by the
weight of the hyperloop which connects all the spins: $\langle
\tau_1\ldots\tau_m \rangle_{ _{\textrm{TPM}}} \sim
    [\tanh(\beta)]^{N^{d_{\textrm{H}}/2}}$.  We can therefore conclude
    by noticing that for every $m$ one has \be
    \lim_{N\rightarrow\infty} \binom{M_L}{m} \langle
    \tau_1\ldots\tau_m \rangle_{ _{\textrm{TPM}}} \sim N^m
        [\tanh(\beta)]^{N^{d_{\textrm{H}}/2}} = 0.  \ee

\subsubsection{{\bf Estimate of $\overline{\mathcal{Z}^2}$}}
\label{A1ab}

Making use of the argument discussed in App.~({\ref{A1aa}), we show
  here that the partition function of the modified TPM of
  Sec.~\ref{S1a} is self-averaging, namely that we have
  $\overline{\mathcal{Z}}^2=\overline{\mathcal{Z}^2}$. Let us first
  write the expression of the high-temperature series of
  $\mathcal{Z}^2$ in a convenient way and then take the average over
  the disorder. We have that, for a given instance of the disorder,
  $\mathcal{Z}^2$ reads as:
\begin{widetext}
\bea
\mathcal{Z}^2 &=& \sum_{\boldsymbol \sigma,{\bf s}} e^{-\beta H_{\textrm{TPM}}[\boldsymbol \sigma]-\beta H_{\textrm{TPM}}[{\bf s}] + \beta \sum_{r=1}^{M_L} \sigma_{i(r)}\sigma_{j(r)}\sigma_{k(r)}+s_{i(r)}s_{j(r)}s_{k(r)}} \nonumber \\
&=& 2^{2N}[\cosh(\beta)]^{2M_L+2N} \left\langle \prod_r (1+ \tanh(\beta)[\sigma_{i(r)}\sigma_{j(r)}\sigma_{k(r)}+s_{i(r)}s_{j(r)}s_{k(r)}]+\tanh(\beta)^2 \sigma_{i(r)}\sigma_{j(r)}\sigma_{k(r)}s_{i(r)}s_{j(r)}s_{k(r)}])\right\rangle_{\boldsymbol \sigma,{\bf s}} \nonumber \\
\label{eq:high-temperature-Z2}
\eea 
where we have introduced the average: 
\be
\langle ~\rangle_{\boldsymbol   \sigma,{\bf s}} = \mathcal{Z}_{\textrm{TPM}}^{-2}\sum_{\boldsymbol \sigma,{\bf s}}e^{-\beta(H_{\textrm{TPM}}[\boldsymbol\sigma]+H_{\textrm{TPM}}[{\bf s}])}. 
\ee
It is then useful for what follows to define also:
\bea
\langle ~\rangle_{\boldsymbol   \sigma} &=& \mathcal{Z}_{\textrm{TPM}}^{-1}\sum_{\boldsymbol \sigma}e^{-\beta H_{\textrm{TPM}}[\boldsymbol \sigma]} \nonumber \\ 
\langle ~\rangle_{\bf s} &=&\mathcal{Z}_{\textrm{TPM}}^{-1}\sum_{\bf s}e^{-\beta H_{\textrm{TPM}}[{\bf s}]}
\eea
Now, in order to lighten the notation, is worthing to use the
plaquette variables $\tau_r=\sigma_{i(r)}\sigma_{j(r)}\sigma_{k(r)}$
and $t_r=s_{i(r)}s_{j(r)}s_{k(r)}$ and use the symbol $g_r$ to
represent the polynomial expression in
Eq.(\ref{eq:high-temperature-Z2}) corresponding to the plaquette ``$r$'': \be g_r =
\tanh(\beta)[\tau_r+t_r]+\tanh(\beta)^2 \tau_r\sigma_r, \ee so that
we can write the high-temperature series of $\mathcal{Z}^2$ reads as \be
2^{2N}[\cosh(\beta)]^{2M_L+2N} \left( 1+\sum_r \langle g_r \rangle +
\sum_{r,p} \langle g_r g_p \rangle + \ldots \right).
\label{eq:high-temperature-Z2-g}
\ee
The average of the expression in Eq.(\ref{eq:high-temperature-Z2-g}) over the disorder, which is represented by all the possible ways to 
choose the spins in each of the random plaquettes, is particularly simple and yields
\be
\overline{\mathcal{Z}^2} = 2^{2N} [\cosh(\beta)]^{2M_L} \left( 1 + \sum_{m=1}^{M_L} \binom{M_L}{m} \langle g_1\ldots g_m\rangle_{\boldsymbol   \sigma,{\bf s}}  \right) 
\ee
so that, since in the limit $N\rightarrow\infty$ at fixed $m$ we can write $\binom{M_L}{m} \sim M_L^m =(\alpha_L N)^m$, we have
\be
2^{2N} [\cosh(\beta)]^{2M_L}  ~\leq ~\lim_{N\rightarrow\infty}\overline{\mathcal{Z}^2}~\leq~ 2^{2N} [\cosh(\beta)]^{2M_L} \left( 1 + \sum_{m=1}^{\alpha_L N} (\alpha N)^m \langle g_1\ldots g_m\rangle_{\boldsymbol   \sigma,{\bf s}}  \right)
\label{eq:high-temperature-Z2-g-compact}
\ee
The multiplaquette correlations in Eq.(\ref{eq:high-temperature-Z2-g-compact}) reads in turn
\be
\langle g_1\ldots g_m\rangle = \sum_{k=0}^{m} \binom{m}{k} \left\langle \prod_{i=1}^{m-k}[\tau_i+t_i] \prod_{j=m-k+1}^{m} t_j\tau_j\right\rangle_{\boldsymbol \sigma,{\bf s}}.  
\label{eq:multispin-binomial}
\ee In Eq.(\ref{eq:multispin-binomial}) the lowest degree correlations
are those obtained taking the index $k=0$, namely are of the kind
$\langle \tau_1 \ldots \tau_m \rangle_{\boldsymbol \sigma}$ or
$\langle \tau_1 \ldots \tau_{m-k} \rangle_{\boldsymbol \sigma} \langle
t_{m-k+1} \ldots t_m \rangle_{\bf s} $. The terms on the right hand
side of Eq.~(\ref{eq:multispin-binomial}) where a single correlation
function appears are vanishing in the thermodynamic limit due to the
same argument of App.~(\ref{A1aa}).  With the same kind of arguments
one can show that even the terms $\langle \tau_1 \ldots \tau_m
\rangle_{\boldsymbol \sigma} \langle t_1 \ldots t_m \rangle_{\bf s}$
decay to zero in the thermodynamic limit fast enough to compensate the
combinatorial prefactors, yielding finally the desired result \be
\lim_{N\rightarrow\infty} (\alpha N)^m \langle g_1\ldots
g_m\rangle_{\boldsymbol \sigma,{\bf s}} = 0. \label{eq:last-multiplaquettes} \ee From
Eq.~(\ref{eq:high-temperature-Z2-g-compact}) and Eq.~(\ref{eq:last-multiplaquettes}) it
follows finally \be \overline{\mathcal{Z}^2} =
\overline{\mathcal{Z}}^2 = 2^{2N} [\cosh(\beta)]^{2M_L+2N} \ee
\end{widetext}

\subsection{Additional plaquettes on a regular sublattice: smallest \emph{hyperloop} via simulated annealing}
\label{A1a0}

In Sec.~\ref{S1b} of the paper we mentioned a simulated annealing
method to find the smallest hyperloop in the high-temperature
expansion of Eq.~(\ref{eq:ZGEN}). In order to discuss this method
let us first rewrite the partition function as
\bea && \mZ = \sum_{\lbrace \sigma
  \rbrace}\exp\left[\beta\sum_p\sigma_{i_p}\sigma_{j_p}\sigma_{k_p}\right]
\nonumber \\ &=& \left[\cosh(\beta)\right]^N \sum_{\sigma_1,\ldots,\sigma_N}
\prod_p\left(1+\tanh(\beta)\sigma_{i_p}\sigma_{j_p}\sigma_{k_p}\right)
\nonumber \\ &=& \left[\cosh(\beta)\right]^N
\sum_{\sigma_1,\ldots,\sigma_N} \prod_p \sum_{n_p=0,1}
\left[\tanh(\beta)\sigma_{i_p}\sigma_{j_p}\sigma_{k_p}\right]^{n_p},\nonumber \\ \label{plaqdual0}\eea
where the index $p$ runs over all the plaquettes of the system, both
the plaquettes of the original TPM and the additional plaquettes of
the auxiliary sublattice introduced in Sec.~\ref{S1b}. In the last row
of Eq.~(\ref{plaqdual0}) is convenient to explicitly write the
product over the spins
\bea \mathcal{Z} &=& \left[\cosh(\beta)\right]^N
\sum_{\sigma_1,\ldots,\sigma_N} \sum_{\lbrace n \rbrace} \bigg\lbrace \prod_p
\left[\tanh(\beta)\right]^{n_p} \cdot \nonumber \\ &\cdot& \prod_i \sigma_i^{\left(\sum_{p\in
    \partial i}n_p\right)_{\textrm{mod}2}} \bigg\rbrace \nonumber \\ &=&
\left[\cosh(\beta)\right]^N \sum_{\lbrace n \rbrace} \bigg\lbrace\prod_p
\left[\tanh(\beta)\right]^{n_p} \cdot \nonumber \\ &\cdot& \prod_i \left[ 1 +
  (-1)^{\left(\sum_{p\in \partial i}n_p\right)_{\textrm{mod}2}}\right]\bigg\rbrace,\nonumber \\
\label{eq:plaqdual}
\eea 
where now the index $p$ labels the plaquettes around a given spin,
$p\in \partial i$. The sum over variables $\lbrace n_i
\rbrace_{i=1,\ldots,M}$ appearing in the last rows of
Eq.~(\ref{eq:plaqdual}) represents the sum over all possible collections
of plaquettes, either forming or not an hyperloop. Within a certain
collection of plaquettes the one labeled with $p$ is present when
$n_p=1$ and absent when $n_p=0$. Let us stress that each collection of
plaquettes which does not form an hyperloop, i.e. an assignment of the
variables $n_p$ such that, at least for one $i$, we have $(
\sum_{p\in\partial i} n_p )_{\textrm{mod}2}=1$, does not contribute to
the sum in Eq.~(\ref{eq:plaqdual}). That is why, in order to seek
non-trivial terms of $\mathcal{Z}$, we need to find hyperloops.  A
hyperloop correspond therefore to a choice of $\lbrace n_p \rbrace$
such that for each spin we have \be \left(\sum_{p\in \partial
  i}n_p\right)_{\textrm{mod}2}=0\ee Therefore, in order to find
hyperloops, one can look for the ground states of the dual model
defined by the following energy function: \be \mH_{\textrm{dual}}(N) =
\sum_{i=1}^{N} \sum_{p\in \partial i}n_p.
\label{eq:dualH}
\ee The simulated annealing method comes at this stage as the most
natural one to seek for the ground states of the Hamiltonian in
Eq.~(\ref{eq:dualH}). One introduces an effective inverse temperature
parameter $\beta_{\textrm{eff}} = T_{\textrm{eff}}^{-1}$ and then
samples configurations according to the Boltzmann measure
$\exp(-\beta_{\textrm{eff}} \mH_{\textrm{dual}}(N))$, while slowly
decreasing $T_{\textrm{eff}}$, until a configuration with $E=0$ is
found. In order to find the smallest hyperloop we have realized this
simulated annealing search varying the size $N$ of a TPM with an
auxiliary sublattice made of side $3$ triangles and with open boundary
conditions. Varying $N$ we considered always lattices with a side that
was a multiple of the incommensurate sublattice cell side, namely we
considered triangular lattices of rhomboidal shape and side $L =
6,9,12,\ldots$. The result of our study is that the smallest hyperloop
appears in lattice of side $L=12$ made of $L^2$ plaquettes of the TPM
and $4^2$ plaquettes of the incommensurate sublattice. Such a
hyperloop, which is represented in Fig~\ref{fig:smallest-diagram}, is
made by $54$ plaquettes of the TPM model and $10$ plaquettes of the
incommensurate sublattice. Let us notice that, as argued
  above in App.~(\ref{A1aa}), the hyperloop we have found is symmetric
  with respect of the three symmetry axes of the lattice.

\section{Leaf-removal exact solution for the Random-Diluted TPM on random graph}
\label{A2}

For a random regular graph the action of the ``leaf-removal''
algorithm is represented in the thermodynamic limit by an infinite set
of differential equations~\cite{mrz03} for the connectivities of
spins, $n_\ell(t)$, where $t$ is the reduced time $t=n/N$, with $N$
the total number of spins in the system and $n$ the number of
iterations of the leaf-removal algorithm. Because the maximum possible
value taken by $n$ is $N$ the reduced time is in the interval $\left[
  0,1 \right]$.
Our Random Diluted-TPM differs from the XOR-SAT because for the latter
the initial connectivity $n_\ell(0)$ is Poissonian, while in our case
is the mixed Poissonian/binomial distribution of Eq.~(\ref{eq3}) in
Sec.~\ref{S2} of the main text. The probability distribution in
Eq.~(\ref{eq3}) is the only ingredient of the original Random
Diluted-TPM left when the model is studied on the random regular
graph.  While the behaviour of $n_\ell(t;\alpha_s,\alpha_L)$ can be
studied analytically on the random graph, for the original RD-TPM
model this can be studied only numerically.
The differential equations for the evolution of the connectivity under leaf-removal are
\begin{equation}
\dot{n}_\ell(t) = -\delta_{\ell 1} +\delta_{\ell 0}+\frac{2}{3(\alpha_s+\alpha_L-t)}\left[ (\ell+1) n_{\ell+1}(t)-\ell n_{\ell}(t)\right],
\end{equation} 
where $b(t)=(1-t/(\alpha_s+\alpha_L))^{1/3}$. 
The general solution for $\ell \ge 2$ is:
\begin{equation}
{n}_\ell(t) = b^{2\ell}(t) \sum_{k=0}^\infty n_{\ell+k}(0) \prod_{\rho=0}^{k} (\ell+\rho) \sum_{r=0}^k \frac{[-b^2(t)]^r}{r!} \frac{1}{(k-r)!}.
\label{A2:eq2}
\end{equation}
By assuming the distribution in Eq.~(\ref{eq3}) as initial condition $n_{\ell+k}(0)$ and plugging it into Eq.~(\ref{A2:eq2})
one finds for $\ell\geq 2$:
\begin{equation}
{n}_\ell(t) = \frac{[3 \alpha_L b^2(t)]^\ell}{\ell!} \frac{e^{-3 \alpha_L b^2(t)}}{\mathcal{N}} \sum_{r=0}^3 \mathcal{C}_{r,\ell} B_r(g(t)),  
\end{equation}
where $B_r(x)$ is the Bell polynomial of order $r$ in the variable $x$, $g(t)=3\alpha_L (1-b^2(t))$
and the coefficients $\mathcal{C}_{r,\ell}$ are:
\begin{eqnarray}
\mathcal{C}_{0,\ell} &=& \rho_0 + \ell \frac{\rho_1}{3\alpha_L} + \ell (\ell-1) \frac{\rho_2}{(3\alpha_L)^2} + \ell (\ell-1)(\ell-2) \frac{\rho_3}{(3\alpha_L)^3} \nonumber \\
\mathcal{C}_{1,\ell} &=& \frac{\rho_1}{3\alpha_L} + (2\ell-1) \frac{\rho_2}{(3\alpha_L)^2} + (3\ell^2-6\ell+2) \frac{\rho_3}{(3\alpha_L)^3} \nonumber \\
\mathcal{C}_{2,\ell} &=& \frac{\rho_2}{(3\alpha_L)^2} + 3 (\ell-1)\frac{\rho_3}{(3\alpha_L)^3} \nonumber \\
\mathcal{C}_{3,\ell} &=& \frac{\rho_3}{(3\alpha_L)^3}. \nonumber \\
\end{eqnarray} 
The probability that a certain spin is a leaf at the iteration $t$, $n_1(t)$, is then:
\begin{equation}
n_1(t) = b^2(t) \left[ n_1(0)+\int_0^t ds \left(\frac{4 n_2(s)}{3(\alpha_s+\alpha_L-s)}-1\right)\frac{1}{b^2(s)}\right].
\end{equation}
When the probability to find a leaf vanishes before all the spins are
eliminated, i.e. when $n_1(t)=0$ with $t<1$, the system has a finite
core.  On the contrary when $n_1(t)=0$ only at $t=0$ there is no
core. Studying the behaviour of $n_1(t)$ at different values of
$\alpha_s$ and $\alpha_L$ it is then possible to locate in the
parameter space the line
$(\alpha_s^{\textrm{core}},\alpha_L^{\textrm{core}})$.  
In order to find the SAT/UNSAT transition one needs then to calculate the number of spins 
and plaquettes in the core, when it is present. The concentration of spins in the core reads
\begin{equation}
n_c(t)=\sum_{\ell=2}^{\infty} n_\ell(t),
\label{A2:eq3}
\end{equation}
and by plugging the definition of $n_\ell(t)$ from Eq.~(\ref{A2:eq2})
into Eq.~(\ref{A2:eq3}) one gets:
\begin{equation}
n_c(t)=\frac{e^{-g(t)}}{\mathcal{N}}\left[ -\mathcal{K}_0(t) + \sum_{r=0}^3 \mathcal{K}_r(t) (-g(t)+e^{g(t)} B_r(g(t))) \right],
\end{equation} 

where we have defined $g(t)=3\alpha_L b^2(t)$ and the coefficients $\mathcal{K}_r(t)$
read as: 
\bea
  \mathcal{K}_0(t) &=& \rho_0 + f(t) \left[\frac{\rho_1}{3\alpha_L}-\frac{\rho_2}{(3\alpha_L)^2}+2\frac{\rho_3}{(3\alpha_L)^3} \right] \nonumber \\
  &+& \left( f(t)+f^2(t)\right) \left[ \frac{\rho_2}{(3\alpha_L)^2} - 3 \frac{\rho_3}{(3\alpha_L)^3} \right] + \nonumber \\
  &+& \left( f(t)+3 f^2(t)+f^3(t) \right) \frac{\rho_3}{(3\alpha_L)^3} \nonumber 
\eea
\bea
  \mathcal{K}_1(t) &=&   \frac{\rho_1}{3\alpha_L}-\frac{\rho_2}{(3\alpha_L)^2}+2\frac{\rho_3}{(3\alpha_L)^3} +\nonumber \\
  && + 2 f(t) \left[ \frac{\rho_2}{(3\alpha_L)^2} - 3 \frac{\rho_3}{(3\alpha_L)^3} \right] \nonumber \\ 
  && + \left( f(t)+f^2(t)\right) 3 \frac{\rho_3}{(3\alpha_L)^3} \nonumber 
\eea
\bea
  \mathcal{K}_2(t) &=&  \left[ \frac{\rho_2}{(3\alpha_L)^2} - 3 \frac{\rho_3}{(3\alpha_L)^3} \right] + 3 \frac{\rho_3}{(3\alpha_L)^3}  f(t)\nonumber \\
  \mathcal{K}_3(t) &=&  \frac{\rho_3}{(3\alpha_L)^3} \nonumber \\
\eea 
The number of plaquettes left in the system at time $t$ in the
leaf-removal algorithm is then simply provided by
$m_c(t)=\alpha_s+\alpha_L-t$.  For each point $(\alpha_s,\alpha_L)$ in
the parameter space one must look for the time $t^*$ at which the
leaf-removal algorithm stops and then, if $n_c(t^*)>0$, consider the
ratio $\gamma=m_c(t^*)/n_c(t^*)$: the static transition line is
identified by $\gamma=1$.

\section{Leaf removal and partial traces}
\label{A2b}

Let us justify here some of the arguments in Sec.~\ref{S2b} of the
main text. We briefly explain why in the case when leaf-removal
eliminates all the spins from the interaction network then also the
partition function can be summed by means of partial traces without
giving rise to any close diagram: this is the case when $\mathcal{Z}$
has the simple expression in Eq.~(\ref{eq:free-partition}) of
Sec.~\ref{S2b} of the main text. The method of partial traces consist
in summing the partition function starting from the open boundaries of
the lattice. The necessary condition for this method to work is that
each step there is at least one spin belonging to a single
plaquette. Let us explain how the iterative summation algorithm works:
\begin{enumerate}
\item Look for a spin which appears in a single plaquette, say
  $\sigma_0$.
\item Sum over values of the spin $\sigma_0$ in the partition function:
\begin{eqnarray}
\mathcal{Z} &=& \sum_{\sigma_{0},\ldots,\sigma_{N}} \exp\left(\beta\sum_{\mu=1}^{M_s+M_L} \sigma_{i(\mu)}\sigma_{j(\mu)}\sigma_{k(\mu)}\right) \nonumber \\ 
&=& \cosh(\beta) \sum_{\sigma_{0},\ldots} \left[ 1+\sigma_{0(\nu)}\sigma_{1(\nu)}\sigma_{2(\nu)} \tanh(\beta)\right]\cdot \nonumber \\ && \cdot 
\exp\left(\beta\sum_{\mu\neq\nu}^{M_s+M_L} \sigma_{i(\mu)}\sigma_{j(\mu)}\sigma_{k(\mu)}\right) \nonumber \\ 
&=& 2 \cosh(\beta) \sum_{\sigma_{1},\ldots,\sigma_{N}} \exp\left(\beta\sum_{\mu\neq\nu}^{M_s+M_L} \sigma_{i(\mu)}\sigma_{j(\mu)}\sigma_{k(\mu)}\right) \nonumber \\
\label{eq:partial-tr}
\end{eqnarray}
\item Check if either $\sigma_{1(\nu)}$ or $\sigma_{2(\nu)}$ in
  Eq.~(\ref{eq:partial-tr}), or both, are again participating to a
  single interaction $\gamma\neq\nu$ after the removal of $\nu$ by
  summation. If for instance $\sigma_{1(\nu)}$ is appearing only in
  $\sigma_{1(\nu)}\sigma_{j(\gamma)}\sigma_{k(\gamma)}$, go back to
  point 2). If both $\sigma_{1(\nu)}$ and $\sigma_{2(\nu)}$ are
  participating to more than one interaction, go back to point 1).
\item If and only if the above recursion can be iterated summing over
  all the spin of the system the partition function is the one of
  Eq.~(\ref{eq:free-partition}).  
\end{enumerate}

From the description of the method of partial traces, it is clear that
it is exactly the iterative scheme of leaf-removal. The situation when
leaf-removal leaves a finite core are the situations where closed
diagram arise in $\mathcal{Z}$, so that $\mathcal{Z}$ cannot be simply
summed by means of partial traces. It is worthing to recall that in
order to use leaf-removal or partial traces one must choose the
correct boundary conditions: this is due to the deterministic nature
of these decimation algorithms. Consider for instance the pure
Triangular Plaquette Model with $N$ spins~\cite{gn00}: the partition
function is exactly known to be $\mathcal{Z} = 2^N
[\cosh(\beta)]^{N}$. Nevertheless, in the case of \emph{periodic}
boundary conditions there are no leaves (no spins appearing in a
single plaquette), so that both the leaf-removal algorithm and the
partial traces iterations cannot be started: it looks like the whole
system is a core of dimension $N$. On the contrary, if one considers
open boundary conditions, one finds from partial traces that
$\mathcal{Z} = 2^N [\cosh\log(\beta)]^{N}$, apart from corrections
negligible in the thermodynamic limit. Open boundary conditions are
therefore the correct choice if one wants to use this kind of
algorithms to study the thermodynamics of the system. Moreover, in a
phase diagram like the one of Fig.~\ref{fig:dyn-scaling}, open and
periodic boundary conditions correspond to the same points, because
they differ for a sub-extensive number of plaquettes.

\section{Belief-propagation equations for the Random-Diluted TPM}
\label{A3}

As mentioned in the main text in Sec.\ref{S3}, in order to represent
the two different kind of plaquettes in the system, it is convenient
to introduce two different cavity fields, $v_{\gamma\rightarrow i}$
and $u_{\gamma\rightarrow i}$, respectively for the ``short'' and
``long'' range plaquettes.  These fields allow one to write the
marginal probability distribution of the spin $i$ when all
interactions around it but $\gamma$ are removed, with $\gamma$
representing respectively a short or long range plaquette:

\begin{equation}
p_v(\sigma_i) = \frac{e^{\beta v_{\gamma\rightarrow i}
    \sigma_i}}{2\cosh(\beta v_{\gamma\rightarrow i} )} \qquad 
p_u(\sigma_i) = \frac{e^{\beta u_{\gamma\rightarrow i}
    \sigma_i}}{2\cosh(\beta u_{\gamma\rightarrow i} )}. 
\end{equation}

In order to write in a clear way the Belief Propagation (BP)
equations, we need to introduce also the couple of cavity fields
$\tilde{v}_{i\rightarrow\gamma}$ and $\tilde{u}_{i\rightarrow\gamma}$,
which represent the effective field on $\sigma_i$ when only the
plaquette $\gamma$ is removed, respectively when $\gamma$ is short and
long range. The belief propagation equations for our model read then: 
\begin{eqnarray} 
\tilde{u}_{j\rightarrow\gamma} &=& \sum_{\beta\in \partial j \setminus \gamma}^{n_L-1} u_{\beta\rightarrow j} + \sum_{\beta\in \partial j \setminus \gamma}^{n_s} v_{\beta\rightarrow j}  \nonumber \\ 
\tilde{v}_{j\rightarrow\gamma} &=& \sum_{\beta\in \partial j \setminus \gamma}^{n_L} u_{\beta\rightarrow j} + \sum_{\beta\in \partial j \setminus \gamma}^{n_s-1} v_{\beta\rightarrow j}  \nonumber \\ 
u_{\gamma\rightarrow i} &=& \frac{1}{\beta} \tanh^{-1}\left(
  \tanh(\beta) \prod_{j \in \partial\gamma\setminus i} \tanh(\beta
  \tilde{u}_{j\rightarrow\gamma}) \right) \nonumber \\
v_{\gamma\rightarrow i} &=& \frac{1}{\beta} \tanh^{-1}\left(
  \tanh(\beta) \prod_{j \in \partial\gamma\setminus i}  \tanh(\beta
  \tilde{v}_{j\rightarrow\gamma}) \right),\nonumber \\
\label{eq:BP}
\end{eqnarray}
where $n_L$ and $n_s$ denotes respectively the number of long and
short range plaquettes attached to each spin $j$. Let enclose the
belief propagation equations in Eq.~(\ref{eq:BP}) in the expression
\begin{equation}
u_{\gamma\rightarrow i} = \mathcal{F}\left(\lbrace u_{\beta\rightarrow j} \rbrace_{ j \in \partial\gamma\setminus i } \right).
\label{eq:F}
\end{equation}
The population dynamics algorithm is realized starting with a
sufficiently large sample of values for each of the two fields $u$ and
$v$, randomly initialized with flat distribution in the interval
$[-1,1]$.  A random sequential update of the values in the two arrays
is realize according to the BP equations in Eq.~(\ref{eq:BP}). The
numbers of long, $n_L$, and short range plaquettes, $n_s$, attached to
each spin and necessary for each iteration step of the algorithm are
random variables extracted according to the distributions of
Eq.~(\ref{eq:dist2}) in the text.
As is clear from the first two lines of Eq.~(\ref{eq:BP}), 
at each iteration step one also needs the \emph{excess degree distributions} $p_{exc}$
and $\rho_{exc}$, defined as follows. If we already know that the long-range plaquette
$\gamma$ is attached to the spin $\sigma_i$, $p_{exc}$ is the probability
that $n_L-1$ other long-range plaquettes are attached to $\sigma_i$. 
The same is true for the definition of $\rho_{exc}$ when we know that 
$\gamma$ is a short range plaquette. 
Such distributions read respectively
\begin{equation}
p_{exc}(n_L-1) = \frac{n_L p(n_L)}{\langle n_L \rangle} \qquad \rho_{exc}(n_s-1) = \frac{n_s \rho(n_s)}{\langle n_s \rangle}
\label{eq:exc-dist}
\end{equation}
On the Bethe lattice our model has both a dynamic phase transition, at
$T_d$, and thermodynamic transition, at $T_K$. While the dynamic
ergodicity breaking at $T_d$ disappears in interaction networks with
with finite loops, the ideal glass transition at $T_K$ may survive.
The dynamical transition temperature $T_d$ correspond to the formation
of an exponentially large number of metastable states separated by
extensive barriers and such that the system, when initialized in one
of this states is trapped within it. In term of the cavity equations
this phenomenon can be recognized by introducing a couple of auxiliary
fields $u_{\sigma=\pm 1}$ (an $v_{\sigma=\pm 1}$) for each type of
cavity field ($u$ and $v$). The cavity fields $u_\sigma$ represent the
value of the field on $\sigma$ conditioned to the knowledge of the
value taken by this spin, either $\sigma=1$ or $\sigma=-1$. If the
populations of $u_{\sigma=1}$ and $u_{\sigma=-1}$, which are
calculated according to the equations below, Eq.~(\ref{eq:1RSBm1-iteration}), at equilibrium are such that the
$\mathcal{P}(u)=\mathcal{P}_{1}(u_1)=\mathcal{P}_{-1}(u _{-1})$, where
$\mathcal{P}$, $\mathcal{P}_{1}$ and $\mathcal{P}_{-1}$ are the
probability distributions respectively of $u$, $u_{-1}$ and $u_{+1}$,
it means that the system is in the simple paramagnetic state. On the
contrary when the two distributions $\mathcal{P}_{1}(u_1)$ and
$\mathcal{P}_{-1}(u _{-1})$ become different from the the distribution
$\mathcal{P}(u)$ of the equilibrium field, it means that the system
``remembers'' the initial condition and the effective field around a
certain spin favors the values taken by such spin in the initial
condition: ergodicity is dynamically broken. The equation to
recursively update the distributions $\mathcal{P}_\sigma(u_\sigma)$ is
the following,

\begin{widetext}

\begin{eqnarray}
\mathcal{P}_\sigma(u_{\sigma}|u) &=& \sum_{\lbrace m_L(i), n_s(i)
  \rbrace} \prod_{i=1}^2 p_{exc} (m_L(i)) \rho (n_s(i)) \int \left[
  \prod_{i=1}^2 \prod_{j}^{m_L(i)} \prod_{k}^{n_s(i)} du^j dv^k
  \mathcal{P}(u^j) \mathcal{P}(v^k)\right]
\delta(u-\mathcal{F}(\lbrace u^j ,v^k \rbrace)) \nonumber \\ &&
\sum_{\sigma_1\sigma_2} 
\frac{e^{\beta \sigma\sigma_1\sigma_2}}{\mathcal{Z}(\lbrace u^j,v^k \rbrace)}
\prod_{i=1}^2  \prod_{j}^{m_L(i)} \prod_{k}^{n_s(i)} \frac{e^{\beta
    u^j \sigma_i}}{2\cosh(\beta u^j)} \frac{e^{\beta v^k
    \sigma_i}}{2\cosh(\beta v^k)} \nonumber \\ 
 &\cdot& \int \left[ \prod_{i=1}^2 \prod_{j}^{m_L(i)}  d
   u_{\sigma_i}^j \mathcal{P}_{\sigma_i}(u_{\sigma_i}^j|u^j)
   \prod_{k}^{n_s(i)}  d v_{\sigma_i}^k \mathcal{P}_{\sigma_i}(v_{\sigma_i}^k|v^k)
\right] \delta(u_\sigma-\mathcal{F}(\lbrace u_{\sigma_i}^j,v_{\sigma_i}^k \rbrace)), 
\label{eq:1RSBm1-iteration}
\end{eqnarray}

\end{widetext}

where $n_s(i)$ is drawn from $\rho_{n}$ in Eq.~(\ref{eq:dist2}) while
$m_L(i)$ is drawn from $\rho_{exc}$ in Eq.~(\ref{eq:exc-dist}).  The
same kind of equations holds for $\mathcal{P}_\sigma(v_{\sigma}|v)$,
just with $n_L(i)$ (drawn from $p_n$ in Eq.~(\ref{eq:dist2}) in place
of $m_L(i)$ and $m_s(i)$ (drawn from $p_{exc}$ in
Eq.~(\ref{eq:exc-dist}) in place of $n_s(i)$.

The iteration step of the population dynamics according to 
Eq.~(\ref{eq:1RSBm1-iteration}) proceed as follows: 

\begin{enumerate}
\item Choose an element to update in the population of fields
  $u_{\sigma}$, which is equivalent to say: choose randomly a spin
  $\sigma$ in the lattice. It is assumed that we are interested in the
  cavity field on $\sigma$ that is obtained by removing all the
  plaquettes but one, say the plaquette $\gamma$. When studying the
  distribution of $u_{\sigma}$ we know that $\gamma$ is a long range
  plaquette.

\item Consider the spins $\sigma_i$ which are interacting with
  $\sigma$ through the plaquette $\gamma$. Extract then the number of
  long-range, $m_L(i)-1$, and short-range, $n_s(i)$, plaquettes
  attached to each of the spins $\sigma_i$.

\item Compute the cavity field $u$ according to the function
  $\mathcal{F}$ in Eq.~(\ref{eq:F}) from the cavity fields $\lbrace
  u^j,v^k\rbrace$.

\item Choose the values of spins $\sigma_i$ according to equilibrium
  measure for a given value of $\sigma$, namely with probability
  $p(\sigma_1,\sigma_2|\sigma) =\frac{e^{\beta
      \sigma\sigma_1\sigma_2}}{\mathcal{Z}(\lbrace u^j,v^k \rbrace)}
  \prod_{i=1}^2 \prod_{j}^{m_L(i)} \prod_{k}^{n_s(i)} \frac{e^{\beta
      u^j \sigma_i}}{2\cosh(\beta u^j)} \frac{e^{\beta v^k
      \sigma_i}}{2\cosh(\beta v^k)} $.
\item For each cavity field $u^j$ ($v^k$) consider the attached
  $u_{\sigma_i}^j$ ($v_{\sigma_i}^k$).
\item From the set of $\lbrace u_{\sigma_i}^j, v_{\sigma_i}^k \rbrace$
  update the field $u_{\sigma}$ ($v_{\sigma}$) according to
  $\mathcal{F}$.
\end{enumerate}

Once the population dynamic algorithm is converged and the stationary
probability distributions also for the conditioned cavity fields
$u_\sigma$ and $v_\sigma$ is know, one can calculate from it the
free-energy within a single metastable state according to the
following formula. 

\begin{widetext} 

\begin{equation}
f_{\textrm{meta}} = \alpha_{L} \langle f_{L}(u,v, u_\sigma,v_\sigma) \rangle +
\alpha_{s} \langle f_{s}(u,v,u_\sigma,v_\sigma) \rangle 
- \sum_{n} \mathcal{Q}_n(\alpha_s,\alpha_L) (n-1) \langle f_{\sigma}^{(n)}(u,v,u_\sigma,v_\sigma)\rangle, 
\end{equation} 

where we have called here $\mathcal{Q}_n(\alpha_s,\alpha_L)$ the mixed
Poisson/binomial probability distribution of spin connectivity defined
in Eq.~(\ref{eq3}) as $n_\ell(\alpha_s,\alpha_L)$. The free-energy per
plaquette reads, in the case of a long range plaquette, as

\begin{eqnarray}
\langle f_{L}(u,v, u_\sigma,v_\sigma) \rangle &=& 
-\frac{1}{\beta}\sum_{\lbrace \sigma_i \rbrace,
  i\in \partial\triangle} ~\sum_{\lbrace m_L(i), n_s(i)\rbrace } ~
\prod_{i=1}^3 p_{exc}(m_L(i))\rho(n_s(i)) \int \left[ \prod_{i=1}^2
  \prod_{j}^{m_L(i)} \prod_{k}^{n_s(i)} du^j dv^k \mathcal{P}(u^j)
  \mathcal{P}(v^k)\right] \nonumber \\ && \frac{e^{\beta \prod_{i=1}^3 \sigma_i}}{\mathcal{Z}(\lbrace u^j,v^k \rbrace)} 
\prod_{i=1}^3 \prod_{j=1}^{m_L(i)} \prod_{k=1}^{n_s(i)} \frac{e^{\beta u^j \sigma_i} }{2\cosh(\beta u^j)} \frac{e^{\beta v^k \sigma_i} }{2\cosh(\beta v^k)} \nonumber \\
&& \int \left[ \prod_{i=1}^2 \prod_{j}^{m_L(i)}  d u_{\sigma_i}^j
  \mathcal{P}_{\sigma_i}(u_{\sigma_i}^j|u^j) \prod_{k}^{n_s(i)}  d v_{\sigma_i}^k \mathcal{P}_{\sigma_i}(v_{\sigma_i}^k|v^k)
\right] 
\log \mathcal{Z}_{\triangle}(\lbrace u_{\sigma_i}^j ,v_{\sigma_i}^k \rbrace).
\label{eq:free-ene-plaq}
\end{eqnarray}
The expression in Eq.~(\ref{eq:free-ene-plaq}) that has to be consistently modified for a short range
plaquette. The free-energy per spin is then 
\begin{eqnarray}
&& \langle f_{\sigma}^{(n)}(u,u_1,u_{-1})\rangle = \sum_{\lbrace \sigma \rbrace} \int \left[\prod_{j}^{m_L} \prod_{k}^{n_s} du^j dv^k \mathcal{P}(u^j)
  \mathcal{P}(v^k))\right] 
\frac{1}{\mathcal{Z}(\lbrace u^j,v^k \rbrace)} \prod_{j=1}^{m_L} \prod_{k=1}^{n_s}
\frac{e^{\beta u^i \sigma}}{2\cosh(\beta u^i)} \frac{e^{\beta v^k \sigma}}{2\cosh(\beta v^k)} \nonumber \\
&& \int \left[ \prod_{j}^{m_L(i)}  d u_{\sigma_i}^j
  \mathcal{P}_{\sigma_i}(u_{\sigma_i}^j|u^j) \prod_{k}^{n_s(i)}  d v_{\sigma_i}^k \mathcal{P}_{\sigma_i}(v_{\sigma_i}^k|v^k) \right] \log 
\mathcal{Z}_{\sigma}(\lbrace u_{\sigma}^j, v_{\sigma}^k \rbrace). 
\label{eq:free-ene-spin}
\nonumber \\ 
\end{eqnarray}
The two partition functions $\mathcal{Z}_\triangle$ and
$\mathcal{Z}_\sigma$ appearing respectively in
Eq.~(\ref{eq:free-ene-plaq}) and Eq.~(\ref{eq:free-ene-spin}) are define
as
\begin{eqnarray} 
\mathcal{Z}_{\triangle}(\lbrace u_{\sigma_i}^j, v_{\sigma_i}^k \rbrace) &=& \sum_{\lbrace \sigma_i \rbrace}
e^{\beta \prod_{i=1}^3 \sigma_i} \prod_{i=1}^3 
\prod_{j=1}^{m_L(i)} \prod_{k=1}^{n_s(i)} \frac{e^{\beta u^j \sigma_i}
}{2\cosh(\beta u^j)} \frac{e^{\beta v^k \sigma_i} }{2\cosh(\beta v^k)}
\nonumber \\
\mathcal{Z}_{\sigma}(\lbrace u_{\sigma_i}^j, v_{\sigma_i}^k
\rbrace) &=& \sum_\sigma \prod_{j=1}^{m_L} \prod_{k=1}^{n_s} \frac{e^{\beta u^j \sigma}
}{2\cosh(\beta u^j)} \frac{e^{\beta v^k \sigma} }{2\cosh(\beta v^k)}
\end{eqnarray} 
The free energy in the paramagnetic phase is the same as in the
high-temperature expansion, 
\begin{equation} 
f_{\textrm {para}} = -\beta^{-1}\log(2)-\beta^{-1}(\alpha_s+\alpha_L)\log\cosh(\beta).
\end{equation}
The configurational entropy can be finally obtained as:
\begin{equation}
\Sigma = \beta(f_{\textrm{meta}} - f_{\textrm {para}})
\end{equation}
\end{widetext}
\bibliographystyle{apsrev}
\end{document}